\documentclass[a4paper,fleqn]{cas-dc}

\usepackage[utf8]{inputenc}
\usepackage[english]{babel}

\usepackage{amssymb}
\usepackage[english]{babel}
\usepackage{hyperref}
\usepackage{graphicx}
\usepackage{lipsum}
\usepackage{textgreek}
\usepackage{lscape}
\usepackage{amsmath}
\usepackage{multirow}
\usepackage[figuresright]{rotating}
\usepackage{lineno}
\usepackage{algorithm} 
\usepackage{algpseudocode} 
\usepackage{comment}
\usepackage{lineno}
\usepackage{nomencl}
\makenomenclature
\usepackage{calc}
\usepackage[T1]{fontenc}
\usepackage{relsize}
\usepackage{subcaption}
\usepackage{soul}
\usepackage[capitalise,nameinlink,noabbrev]{cleveref}
\usepackage{adjustbox}
\usepackage{multirow}

\usepackage[numbers]{natbib}

\begin{document}



\hypersetup{
  citecolor=Violet,
  linkcolor=Red,
  urlcolor=Blue}

\newcolumntype{P}[1]{>{\centering\arraybackslash}p{#1}}

\def\tsc#1{\csdef{#1}{\textsc{\lowercase{#1}}\xspace}}
\tsc{WGM}
\tsc{QE}
\tsc{EP}
\tsc{PMS}
\tsc{BEC}
\tsc{DE}



\let\WriteBookmarks\relax
\def\floatpagepagefraction{1}
\def\textpagefraction{.001}
\shorttitle{Detecting Plant VOCs with Indoor Air Quality Sensors}
\shortauthors{Nabaei et~al.}

     
\title [mode = title]{ Detecting Plant VOC Traces using Indoor Air Quality Sensors}

\author[1]{Seyed Hamidreza Nabaei}\ead{fgx9eq@virginia.edu}
\author[2]{Ryan Lenfant}\ead{rpl8af@virginia.edu}
\author[3]{Viswajith {Govinda Rajan}}\ead{gyx4bw@virginia.edu}
\author[2]{Dong Chen}\ead{dqc4vv@virginia.edu}
\author[4]{Michael P. Timko}\ead{mpt9g@virginia.edu}
\author[2,3]{Bradford Campbell}\ead{bradjc@virginia.edu}
\author[5]{Arsalan Heydarian}\cormark[1]\ead{ah6rx@virginia.gov}

\cortext[1]{Corresponding author.}



\address[1]{Department of Systems and Information Engineering, link lab, University of Virginia, Charlottesville, VA 22904, USA}
\address[2]{Department of Computer Science, University of Virginia, Charlottesville, VA 22904, USA}
\address[3]{Department of Electrical and Computer Engineering, University of Virginia, Charlottesville, VA 22904, USA}
\address[4]{Department of Biology, University of Virginia, Charlottesville, VA 22904, USA}
\address[5]{Department of Civil and Environmental Engineering, University of Virginia, Charlottesville, VA 22904, USA}

\begin{abstract}
In the era of increasing interest in healthy buildings and the advent of smart homes, the significance of sustainable and health-conscious living spaces is paramount. Smart tools, including VOC sensors, play a pivotal role in monitoring indoor air quality. However, effective monitoring and interpreting signals from indoor VOC sources remains a challenge. To address this challenge, a new approach lies in understanding how indoor plants act in response to environmental conditions. Indoor plants produce terpenes, a type of VOC, in response to various abiotic and biotic stressors, such as detection of airborne pathogens, defense against predators, and intense light and temperature. Changing indoor plant VOCs thus offer the potential for monitoring indoor air quality. Existing research on signal interpretation relies heavily on specialized laboratory sensors to monitor plant responses. Our work aimed to leverage readily available commercial sensors to interpret and classify VOC emitted from plants as a potential means for monitoring changes in indoor air that contribute to the well-being and living experience of occupants in smart buildings. Specifically, we quantified the sensitivity of commercial air quality sensors by measuring 16 terpenes in a controlled setting. We identified the most promising terpenes and measured their concentrations in a realistic indoor environment. We investigated the use of physics-based models to map the response of the VOCs, but recognized that these models struggle to match the complexities of real-world environments. We then trained machine learning models to accurately classify terpenes using commercial VOC sensors and identified the optimal location for placing a sensor. To corroborate our findings, we analyze the terpene emissions of a living basil plant and successfully demonstrate the system's ability to detect VOC emissions from the plant. 
%
%
Our findings established a foundation for overcoming the challenges associated with detecting and classifying plant VOC emissions, paving the way for sophisticated plant-based sensors to significantly improve indoor environmental quality in smart buildings of the future.
\end{abstract}

\begin{keywords}
\sep Indoor Air Quality Sensor \sep Plants \sep Terpenes \sep Machine Learning \sep Volatile Organic Compounds~(VOCs)

\end{keywords}

\maketitle

\section{Introduction and Background}
\label{sec:intro}

Being sedentary organisms, plants have evolved remarkable mechanisms to rapidly sense and respond to changes in their surrounding environments 
Plants are naturally attuned to their environments, capable of sensing changes in air quality ~\cite{matthen2024plants, napier2023gene, demey2023sound}, temperature ~\cite{kerbler2023temperature}, presence of specific airborne chemicals or pollutants ~\cite{hong2023plants, matsui2016portion}, and even presence of human ~\cite{de2023can}.
In a study, several indoor plants were used as biomonitors to detect toxic metals (Cd, Cr, Pb) from tobacco smoke ~\cite{ghoma2022using}. Leaf samples were analyzed using inductively coupled plasma-optical emission spectrometry (ICP-OES) to measure metal accumulation. 
Recent studies have demonstrated that plants' sensing mechanism, developed over millions of years, offers invaluable insights and inspiration for addressing modern challenges in indoor environmental monitoring.
Zuniga et al. (2022) introduced the concept of ``smart plants'', referring to plants integrated with CO2, temperature, and humidity sensors. These enhanced plants were used to monitor indoor air quality and occupancy levels, exploring whether they could complement and improve traditional sensing systems. Plant physiological changes, such as photosynthesis rate and stomatal conductance, were observed as they reacted to CO2 fluctuations caused by varying occupancy levels. These biosensors detected CO2 accumulation, indicating human presence and activities, and changes in temperature and humidity that correlated with occupancy and mask use. The data provided real-time insights into indoor environmental conditions, offering a cost-effective and easy-to-deploy alternative to traditional sensor installations ~\cite{zuniga2022smart}.

In another study, researchers augmented plants to detect motion by embedding nanowires in the tissue of the plant ~\cite{sareen2017cyborg}. These nanowires acted as sensors that detected human touch or nearby movement, triggering responses in the plant such as changes in electrical signals or leaf movements. This integration allowed the plant to function as a motion sensor, reacting to physical interaction through shifts in its physiological state. Another study monitored air pollution levels in terrestrial environments, using various plant species, such as lichens and mosses, as bioindicators ~\cite{mulgrew2000biomonitoring}. They analyzed the plants’ ability to accumulate pollutants like heavy metals and ozone to assess air quality. The study aimed to evaluate the effects of air pollution on ecosystems and human health, providing insights into contamination levels and contributing to pollution control efforts.

Plants employ various signaling mechanisms, including the release of VOCs \cite{HEDRICH2016376, chemcomm,riedlmeier2017monoterpenes}, changes in leaf color ~\cite{kohzuma2024analysis}, electrical signals ~\cite{fromm2007electrical}, and other physiological responses ~\cite{madani2019physiological}. In addition, plants can actively identify and respond to specific types of airborne particles, including chemical signals from other plants or threats such as pathogens and pests ~\cite{loreto2022plants}. For example, plants can detect the VOCs released by neighboring plants under attack, prompting them to preemptively activate their own defense mechanisms ~\cite{waterman2024high, loreto2022plants}. 
Terpenes are one class of plant VOCs that are commonly used to respond to external  stimuli, such as predation by insect pests and herbivores or signaling reproductive readiness \cite{riedlmeier2017monoterpenes, bouwmeester2019role}. 
Each unique blend of VOCs carries specific information for predators, pollinators, and neighboring plants~\cite{baldwin2010plant}. 
For example, when attacked by arthropods, a tomato plant can release a complex blend of up to 70 VOCs, including compounds like (E)-β-ocimene, linalool, and allo-ocimene~\cite{miano2022electroantennogram}. 
The release of these terpenes serves two primary functions: (1) to communicate the presence of airborne threats to nearby plants and (2) to act as a defense mechanism, such as attracting natural predators to eliminate pests~\cite{parasiteaid}. 
Although VOC-time profiles can identify the released VOCs, identical dosages do not always produce consistent profiles  ~\cite{yeoman2021estimating}.  Studies indicate that even with controlled timing and dosage, peak concentrations can vary ~\cite{yeoman2021estimating}. However, patterns in VOC fluctuations and terpene emissions may offer valuable insights into how plants respond to specific environmental changes.


Recent breakthroughs in synthetic biology have allowed scientists to enhance the natural capabilities of plants by genetically modifying them to detect and absorb specific pollutants from the air ~\cite{zhang2018greatly, aminian2023synthetic}. These innovations transform plants into active participants in environmental remediation, enabling them to target harmful airborne substances more effectively. Synthetic biology approaches, traditionally used in microbial engineering, are now applied to plants to tackle environmental pollutants, offering new pathways for bioremediation by enhancing their ability to interact with and remove contaminants from the atmosphere ~\cite{aminian2023synthetic, volkov2006plants, jin2020synthetic}. 
This includes modifying plants to sense particular chemical compounds, pollutants, or even biological agents such as bacteria, viruses, or fungal spores ~\cite{maurya2023microbially,  barbinta2024nature, manginell2006viral}. 

Furthermore, researchers have engineered plants to detect specific viral particles by enhancing their natural antiviral mechanisms, such as small RNA-mediated interference, which helps them recognize and degrade viral RNA ~\cite{majumdar2023natural}. This innovation offers a promising new approach for early disease detection in agriculture or controlled indoor environments.

Genetically modified plants have the potential to act as "living sensors" in various environments, both indoors and outdoors ~\cite{WOLKOFF2020113439}. These engineered plants can be designed to monitor air quality, detect harmful chemicals, or signal the presence of pathogens. By introducing specific genetic modifications, plants can be engineered to react to pollutants or toxins in the environment, often through visible changes such as color shifts or fluorescence. This capability allows them to function as bioindicators for detecting pollutants like VOCs, heavy metals, or harmful pathogens, making them a valuable tool for real-time environmental monitoring. With their natural ability to grow and adapt to different surroundings, modified plants offer a sustainable and integrative approach to tracking environmental changes and maintaining ecosystem health ~\cite{WOLKOFF2020113439}.

While plants naturally emit VOCs in varying dosages and time patterns in response to different stimuli, accurately interpreting these emissions requires advanced sensing technologies capable of real-time analysis. Low-cost VOC sensors, such as those used in indoor air quality monitoring, lack the precision needed to identify specific phytochemicals and to quantify their levels accurately. Two widely used methods for plant VOC analysis—gas chromatography and mass spectrometry—offer high-precision results but are typically confined to laboratory settings ~\cite{materic2015methods}.
Electronic noses (E-noses) offer an alternative approach, allowing researchers to capture and analyze VOC emissions from plants with high sensitivity. By measuring the normalized sensor response, reaction time, and VOC dissipation rates, E-noses can accurately distinguish specific VOCs emitted by plants ~\cite{gancarz2019identification}. These systems leverage expensive equipment like high-resolution gas sensors, advanced artificial neural network (ANN) algorithms, and sophisticated data processing to analyze complex odor profiles, yielding detailed insights into plant behavior that are often beyond the reach of commercial sensor systems ~\cite{fitzgerald2017artificial}.
While E-noses are effective at identifying and classifying VOCs ~\cite{rasekh2021performance, delgado2012use, dragonieri2017electronic}, their practical use in natural environments, such as open fields, presents significant challenges. High costs, the need for specialized technicians, and limited coverage areas constrain their accessibility in such conditions ~\cite{pathak2022review, macdougall2022emerging}.
On the other hand, commercial sensors provide the flexibility to continuously monitor VOC concentrations in situ and under naturalistic conditions. However, these sensors face critical limitations when applied to plant-emitted VOC detection, particularly in indoor settings where concentrations can range from 1 to 100 parts per billion (ppb). They often struggle with low sensitivity, poor selectivity, and interference from environmental factors. Most commercial sensors are designed for detecting a narrow range of VOCs, primarily targeting indoor pollutants such as formaldehyde, benzene, toluene, and xylene, which limits their ability to capture the full spectrum of VOCs emitted by plants ~\cite{spinelle2017review}.In other words, while commercial sensors can detect fluctuations in VOC levels triggered by plant-emitted terpenes, they are not designed to differentiate between specific VOC compounds.

Recent studies suggest that analyzing VOC emissions as time-series data and applying machine learning techniques—such as principal component analysis, clustering, support vector machines, and artificial neural networks—can achieve over 80\% accuracy in decoding VOC profiles  ~\cite{spadi2022conventional, cui2018plant}. . This approach may help classify specific plant responses and behaviors, bridging the gap between low-sensitivity commercial sensors and high-precision E-noses
In a study, five machine learning classifiers—Logistic Regression (L1 and L2), Decision Trees, Random Forests, Support Vector Machines (SVM), and K-Nearest Neighbors (KNN)—were used to classify VOCs emitted from bacterial and fungal cultures, collected via solid phase micro-extraction (SPME) and analyzed with Ambient Plasma Ionization Mass Spectrometry ~\cite{arora2022machine}. The Random Forest model performed best in distinguishing between bacteria and fungi and identifying specific bacterial strains. This rapid and non-invasive approach shows promise for pathogen detection in clinical settings.
Thorson et al. demonstrated the effectiveness of machine learning techniques in analyzing data from a low-cost sensor array for air quality monitoring ~\cite{thorson2019using}. By applying regression models, such as random forests, to estimate VOC concentrations, they then utilized classification models, including support vector machines, to identify pollutant sources like mobile emissions and biomass burning. This two-step approach achieved an F1 score of 0.72, highlighting its potential for accurate pollutant source attribution. 
Furthermore, a photoionization detector (PID) was employed to differentiate between two rosemary varieties, Rosmarinus officinalis L. “Prostratus” and “Erectus,” by analyzing their VOC emissions ~\cite{spadi2022conventional}. The PID generated distinct ‘fingerprints’ for each variety, which were subsequently validated using advanced machine learning models, including principal component analysis (PCA), support vector machines (SVM), and artificial neural networks (ANN). These models demonstrated a high classification accuracy of over 80\%, confirming the capability of PID as a tool for classifying  the VOC profiles of aromatic plants.This approach highlights the potential of utilizing simple, cost-effective sensors combined with robust statistical analysis to perform VOC classification in plants. What is needed now is a clear demonstration that such comemrcial sensors can be used to monitor changes the indoor air quality in real time with the sensitivity necessary to be predictive and reactive to changes.

The novelty of this study lies in leveraging commercial sensors and machine learning to detect and classify plant VOCs (specifically terpenes) while optimizing sensor placement for accurate data collection. Additionally, this paper introduces a novel approach to interpreting VOC emissions by analyzing patterns over time, including baselines, peak levels, and signal fluctuations, offering deeper insights into VOC behavior.

This paper seeks to harness the potential of machine learning models in identifying and categorizing VOCs emitted by plants under natural conditions, leveraging time-series data gathered from commercially available VOC sensors to open new doors of sensing in buildings. 
The integration of machine learning techniques further equips us to discern subtle variations in VOC emissions, offering a nuanced and comprehensive understanding of plant behavior and their reactions to environmental stimuli. 
This approach significantly improves our ability to understand the intricate dynamics of plant-environment interactions, providing invaluable insights into the complex signaling and communication processes within the realm of plant biology.

To investigate this, we conducted a series of experiments designed to refine sensor placement and enhance classification accuracy using machine learning. The study’s findings demonstrate the high accuracy of our models, ranging from 90\% to 100\%, in identifying various terpenes. Additionally, our work with basil plants under stress conditions confirmed the sensors’ ability to capture changes in terpene emissions, highlighting the potential of this approach for advanced VOC detection in real-world applications. The results pave the way for more nuanced in-situ plant sensing systems, contributing valuable insights to the field of VOC detection. To provide a valuable resource for future research efforts in VOC detection and related fields, the codes and dataset\footnote{Codes: \url{https://github.com/HamidrezaNabaei/Plant-article}} will be made available following the paper's acceptance.

\begin{figure}[t]
\centering
  \centering
  \includegraphics[width=.9\linewidth, keepaspectratio=true]{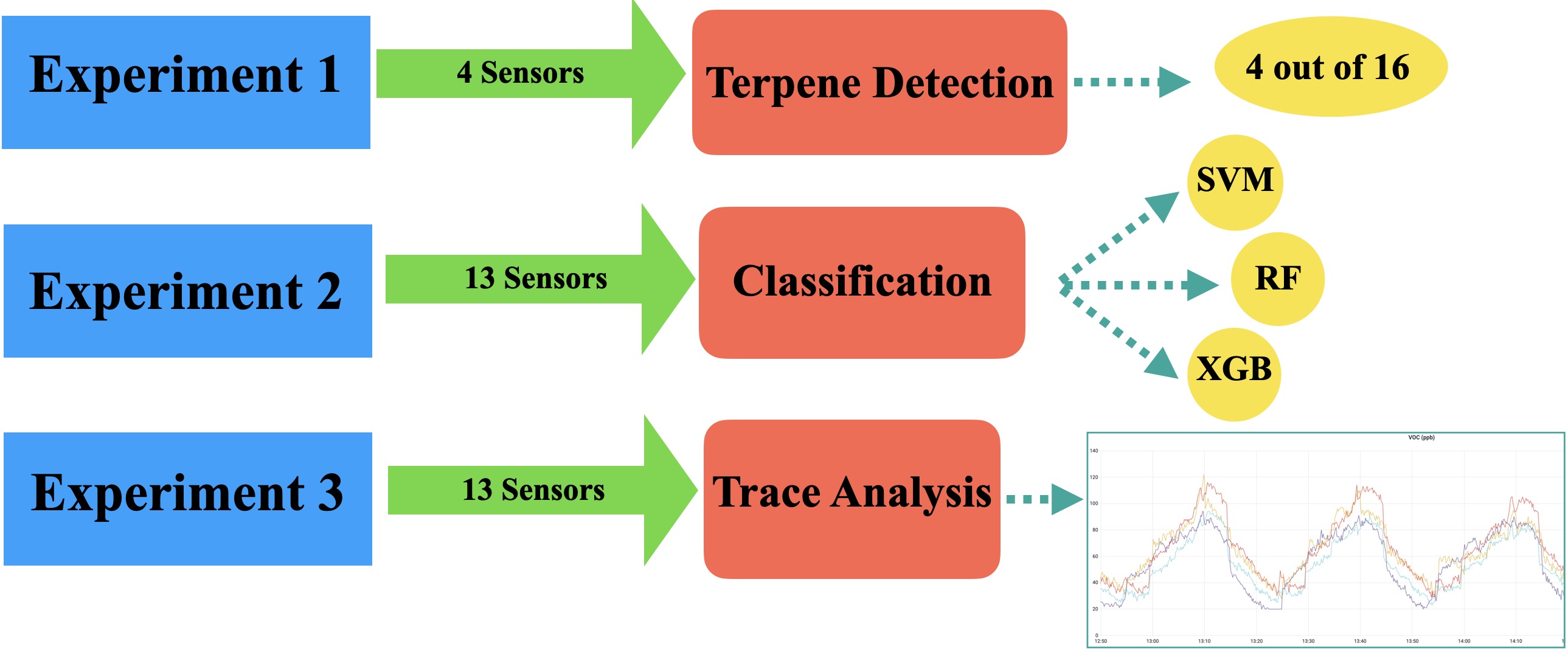}
  \caption{Flowchart showing an overview of the Experiments}
  \label{fig.diagram}
\end{figure}

\section{Methodology}

In this study, our primary objective is to address the following research questions:
\begin{enumerate}
    \item Can commercial sensors detect the presence of terpenes in indoor settings?
    \item How can physics-based and machine learning models be utilized to classify and differentiate terpenes? 
    \item How does the VOC trace differ when plants are in normal condition vs stressed condition? 
\end{enumerate}


To answer these research questions, we conducted three experimental studies as listed below, in each test we used AWAIR Omni sensors (with an accuracy of ±15\% in the working range of  20 - 36,000 ppb), as our commercial sensor, that set up and left in the testing environment for two days to calibrate ~\cite{awairomni}. Figure \ref{fig.diagram} shows an overview of the experiments:



 \begin{itemize}
    \item \textbf{Experiment 1: Commercial sensors} This study is to answer question 1. Various terpenes (including terpene essences and basil plant) were examined within an office setting equipped with a number of Awair Omni sensors to collect VOC data at a 10-second frequency~\cite{awair}. As there are more than 10,000s of different VOCs, commercial sensors do not typically provide specific VOCs and chemicals they can detect; as a result, we selected 16 different commonly found terpenes in plants to determine first, which terpenes can be detected by the Awair Omni sensors. By ditributing four indoor air quality sensors in an office space, we evaluated which of the 16 types of terpenes, the commercial sensors can detect. For this purpose we used commercially available terpene essences. The initial phase involved testing terpenes using 4 sensors positioned around the sample on a planar surface of the table in the room. (Figure \ref{fig: Pilottest}). 
\end{itemize}

\begin{itemize}
    \item \textbf{Experiment 2: Terpene Classification} After finding the terpenes essences that can be sensed by commercial sensors we select 4 most sensible terpene essences to do more experiments on them. To answer question 2, by following the successful completion of experiment 1, which validated the functionality of our sensors in detecting terpene essences and basil plant using identified optimal procedural steps, we proceeded to expand our investigation. The number of sensors was increased to 13, strategically positioned in various locations within the room to capture terpene signals. This expansion aimed to enhance machine learning accuracy by augmenting the dataset and identifying optimal sensor placement (Figure \ref{fig.room.13sensor}).
\end{itemize}

\begin{itemize}
    \item \textbf{Experiment 3: Plant's terpene trace differentiation} Building upon the insights collected from Experiment 2, which encompassed both Experiments 1 and 2, a controlled test box containing a basil plant and a sensor, as illustrated in Figure \ref{fig.Box}, was used to closely examine the plant VOC trace and differentiate between terpene emissions in normal and stressed conditions.
\end{itemize}

\subsection{Experimental procedure}

\label{sec:method:base:awair}
  

\subsubsection{Experiment 1 and 2 Setup}
\begin{figure*}[!t]
    \centering
    \subfloat[\centering Schematic of Room Layout]{{\includegraphics[width=.51\linewidth]{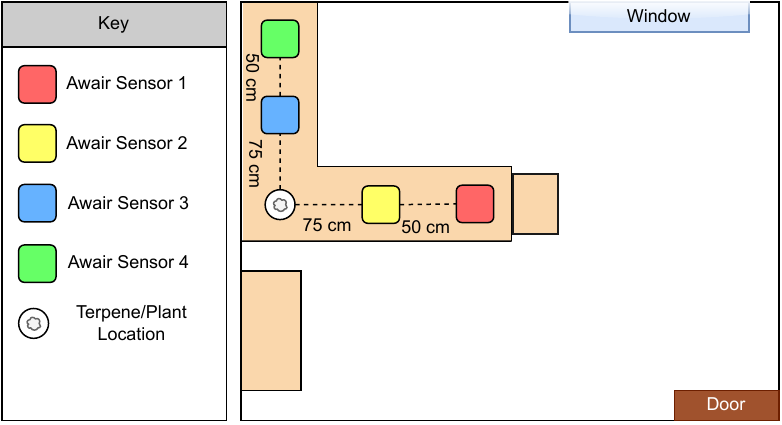} }}%
    \qquad
    \subfloat[\centering Sensor placement - Experiment 1]{{\includegraphics[width=.4\linewidth]{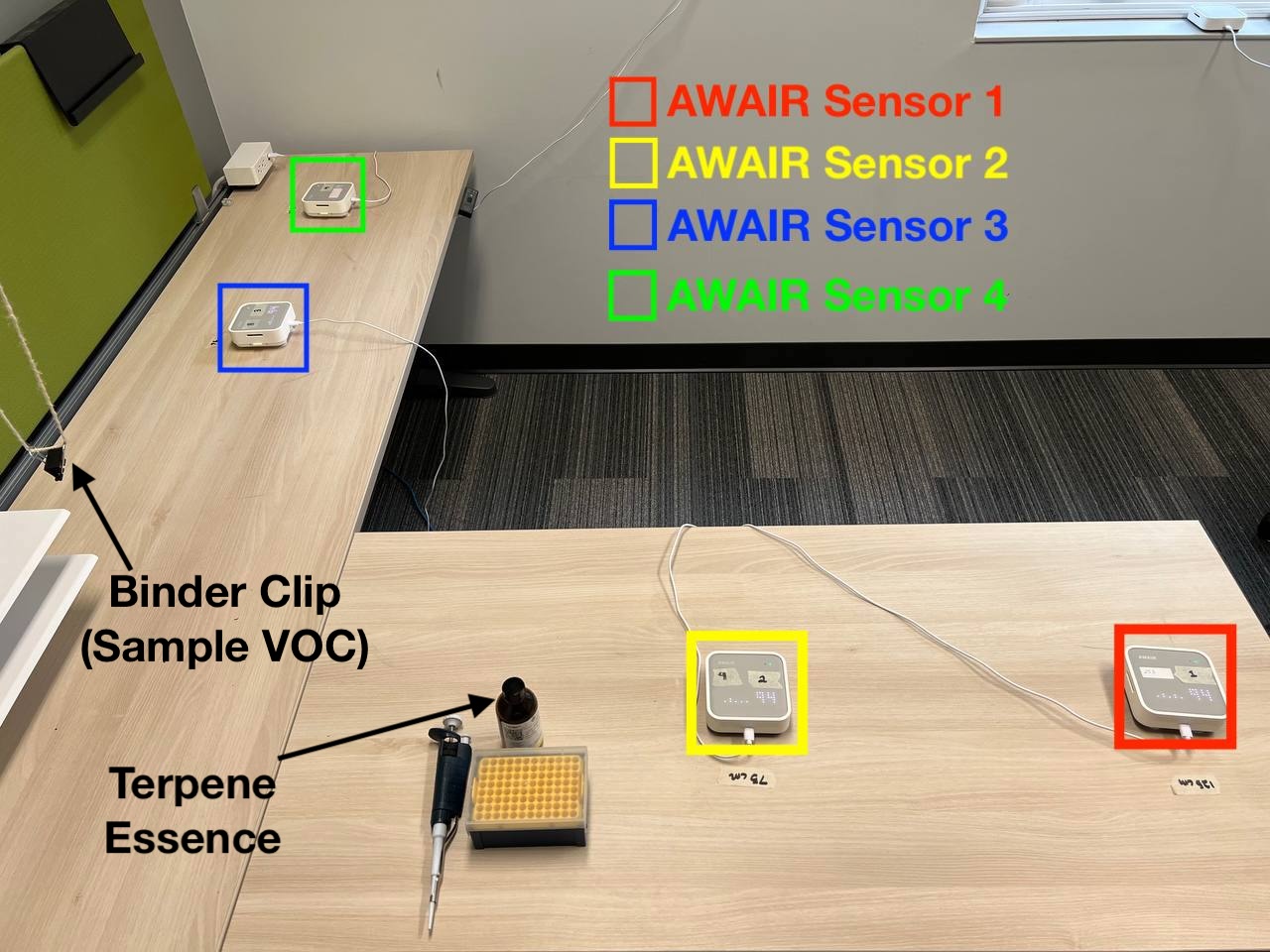}}}
    \caption{This figure shows the layout of the room for the terpene experiment 1. Figure (a) is a representation of the room and
Figure (b) is the actual deployment. 4 sensors were arranged on a table with two of them placed $75~cm$ away and the other two
placed $125~cm$ away from the terpene/plant location respectively}
    \label{fig: Pilottest}
    \vspace{-15pt}
\end{figure*}

\begin{figure}[t]
\centering
  \centering
  \includegraphics[width=1\linewidth, keepaspectratio=true]{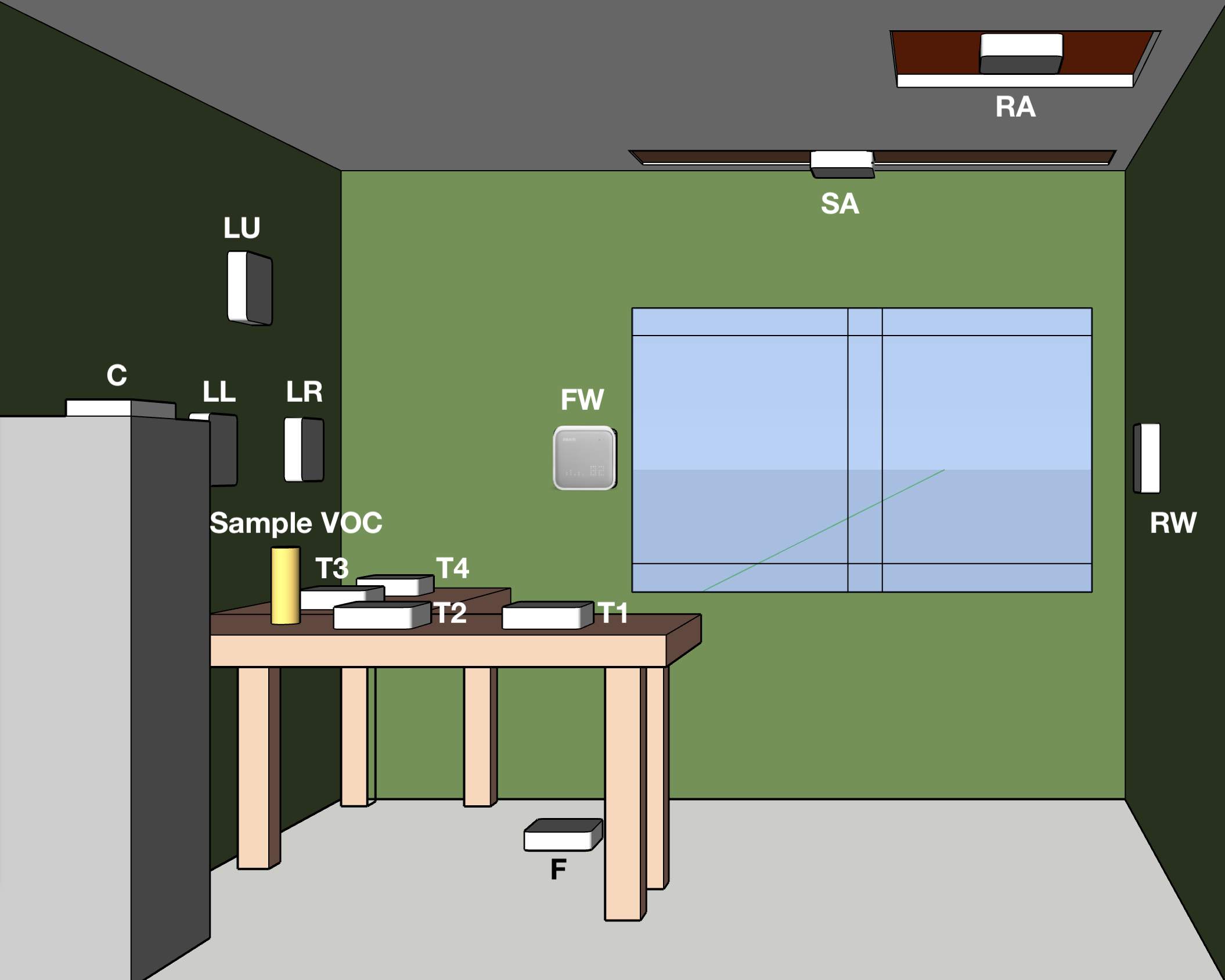}
  \caption{Schematic of sensor placement in Experiment 2 - Testing the VOC terpenes in the room using 13 spatially arranged sensors in the room. The sphere is the placement of the sample and the cubic shapes are the locations of sensors in the room including 6 sensors on the walls, 4 sensors on the table, 1 sensor on the cabinet, 1 sensor on the floor, and 2 sensors on the ceiling, attached to supply and return of the HVAC system. In the picture, sensors are defined with cubic boxes and their symbols from\cref{tab.Sensor_summary}} 
  \label{fig.room.13sensor}
\end{figure}

The experiment was conducted in an office space with a volume of $29.65m^3$ as shown in Figure \ref{fig: Pilottest}, \ref{fig.room.13sensor}. The room is equipped with a controllable HVAC system that utilizes an air handler and Variable Air Volume (VAV) components. All walls are painted and there is a suspended ceiling of panels. There is a carpet on the floor, 2 tables, 1 chair, a cabinet, and a ceiling lamp. 
 The 16 terpenes used in this study and their chemical formulae are listed in~\cref{tab.all_chemicals}. For each individual terpene, a dose of 100 $\mu L$ was applied to a piece of filter paper outside the room.


Immediately after applying the terpene to the filter paper, the sample was placed directly next to the sensor on the desk for 5 minutes. As soon as the samples were placed, the experimenter left the room and no other person was present in the room. The VOC levels for the sensor were monitored during that period and the initial, maximum, and final readings were recorded for the 5-minute period. The difference between the initial and final readings ~($\Delta_{final-initial}$) is used to determine the longevity of the terpene in the room. The difference between the initial and max readings ~($\Delta_{max-initial}$) is used to determine the sensor's sensitivity to terpene. As an outcome of this section, we identified the top-response terpenes to the Awair Omni VOC sensor.

\label{sec.data_collection}
To understand the distinct VOC readings, we collected data for the four most significant plant terpenes'\\($\Delta_{final-initial}$) results from~\cref{tab.all_chemicals}. Previous studies have shown that machine learning can be used to distinguish between different VOC traces~\cite{arora2022machine,gancarz2019identification}. By using the VOC reading of the Awair Omni, we create a repository of VOC profiles for (i) no VOC in the room~(control), (ii) the plant terpenes $\alpha$-Terpinene, Cis-Beta Ocimene, Citral, and D-Limonene, and (iii) basil plants to determine if they are distinguishable from each other.

Experiment 1 was set up in the office space, as mentioned above. To monitor the environment, 4 Awair sensors were set up on a desk, two of which were $75~cm$ away from the terpene sample and the other two were $125~cm$ away from the sample. A binder clip held $30~cm$ above the desk by a string was used to hold the sample in the air. A diagram of the setup is shown in Figure \ref{fig: Pilottest} (a) , while the actual room setup can be seen in Figure \ref{fig: Pilottest} (b). The sensors were left in the room for 1-2 days prior to any testing to calibrate. We also conducted the experiments when the HVAC system was running at normal setting, between 8:00 AM to 6:00 PM during spring and summer which we had the sunlight in the room. 


Before any terpene sample was placed in the room, we made sure each sensor was calibrated to a baseline VOC level ~(<150 ppb). If the sensors did not read at baseline, at the first stage we opened the window and door to clear the VOC levels in the room and then we used an air purifier to cleanse the room until it reached the room's initial VOC baseline. 
To maintain the integrity of the experiments, we took precautions during the entry and exit process. 
After conducting various tests to maintain a constant VOC baseline in the room during entries and exits, we instructed our experimenter to move in and out slowly, taking a cautious approach to avoid the infiltration of external VOCs through the sliding door. 

The experimenter ensured the use of a lab coat and refrained from wearing any cologne, perfume, or other beauty products to minimize the impact on VOC levels in the room. Once all sensors recorded measurements at or below 150 ppb, the experimenter entered the room and applied the intended dosage of the terpene onto the filter paper, subsequently hanging it on the binder clip. The sample was left in the room for 15 minutes, after which the sample was removed from the room and the room settled for 5 minutes. Five minutes later, the door and window were opened to allow the VOCs to dissipate. These steps were replicated for each individual test. The chemicals and dosages used for this study were Cis-Beta Ocimene~(70\%), D-Limonene, Citral, and Alpha-Terpinene at concentrations of 200 $\mu L$ or 100 $\mu L$.

For experiment 1 and 2, the plant is also placed in the same location as the binder clip. The plant is also kept in the room for at least 1-2 days so the Awair sensor would be calibrated based on the presence of the plant. The experimental procedure for the plants was similar to the terpenes where the experimenter waited for the room VOC baseline to reach around 150 ppb. Then, the experimenter entered the room and crushed a leaf of the basil plant. This stressed the plant and made it release more volatile compounds. After 15 minutes the crushed leaves were removed from the room. 
After 5 minutes, the door and window were opened and the air purifier was used to clear out the room VOCs. The only plant used for this study was basil~(\textit{Ocimum basilicum}) serving as our source for terpenes, namely Cis-Beta-Ocimene and D-Limonene.

In Experiment 2 (Figure \ref{fig.room.13sensor}), we replicated the test steps and instructions from Experiment 1 while augmenting the sensor count to 13. This augmentation involved the addition of 9 sensors placed randomly in various locations within the room, including the floor, walls, on top of the cabinet, and at the supply and return points of the air ventilation on the ceiling. Furthermore, we expanded the number of conducted tests to 100, resulting in a comprehensive dataset of 1300 VOC-time data entries. This extensive dataset enhances the precision of terpene categorization in our analysis.
\begin{figure}[t]
\centering
  \centering
  \includegraphics[width=.8\linewidth, keepaspectratio=true]{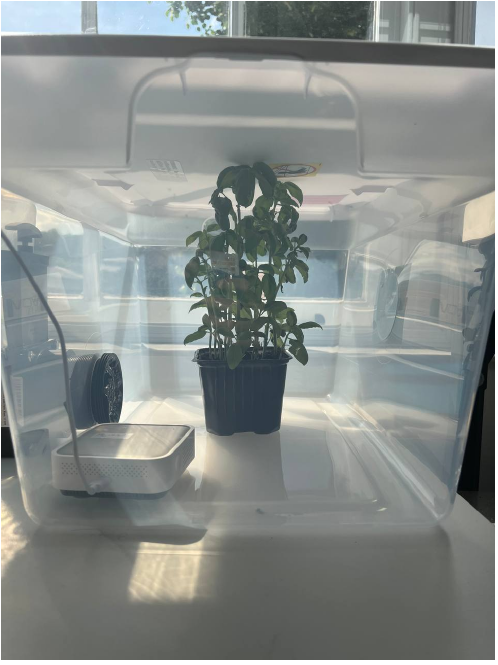}
  \caption{Experiment 3 setup. The basil plant was placed in the center of the box in proximity to a sensor for 30 minutes of testing. The box had no air input or output}
  \label{fig.Box}
\end{figure}

\subsubsection{Experiment 3 Setup}
\label{sec:method:plant}


As it might be difficult to accurately observe the trace of VOC after crushing a leaf, Experiment 3 was designed to closely examine the VOC traces of the basil plant in a controlled boxed environment. For this experiment, we used a 60*40*50 cm polypropylene container (box) ~(\cref{fig.Box}) to test the VOC release of plants, specifically basil. The box was constructed to facilitate controlled tests and provide the plant's terpene emission data. We conducted different tests, including profiling the basil plant itself and crushing one leaf of the plant, respectively. To collect VOC data, we placed a sensor~\cite{awair} inside the box, close to the plant. After many different tests, we found the best testing guideline for finding a signature like VOC detection of plants as a trace of it. This was like during each test, VOC data was collected using the commercial sensor for 30 minutes, and each test was repeated 3-4 times to ensure accuracy and consistency of results. 
We crushed the leaf carefully, while not allowing any external agents to enter the box. After each test, the box was cleaned, and the baseline for each test was measured.
Additionally, the box itself produced a small amount of VOC that counts for different baselines, which was accounted for in the data analysis. 


\subsection{Data Analysis}

Our objective is to differentiate and categorize four different terpenes detected by Awair sensors in indoor environments. Our primary emphasis lies in the effective interpretation of Terpene VOC signals originating from plants. To achieve this, we employ a combination of physics-based models and machine-learning models to classify and distinguish between these different terpenes.

\subsubsection{Physics Model}
\label{phymodel}

Different volatile compounds spread in rooms at different emission rates. To model this difference, this study aims to observe and understand the physics of how VOCs are spread in space. One prior study has used a simple mass-based equation to determine the VOC concentration in the room at a specific time~\cite{hori2013evaluation}. We use this simple mass-based equation to determine the emission rate of the substance in the room. The equation used in the study done by Hori et al.~\cite{hori2013evaluation} can be seen below and assumes the room is a perfect mixing.

\begin{equation}
    C =~(C_{in} - \frac{F}{Q})e^{(-\frac{Q}{V}t)} + \frac{F}{Q}
\end{equation}

This equation solves for C, the concentration in the room~($\mu g/m^3$). $C_{in}$ is the initial concentration~($\mu g/m^3$),  F is the emission rate~($\mu g/h$), Q is the ventilation airflow rate~($m^3/h$), V is the volume of the room~($m^3$) and t is time~($h$). Using this equation, we can solve for the emission rate of the VOC.

\begin{equation}
    F = Q \frac{C -C_{in}e^{(-\frac{Q}{V}t)}}{1-e^{(-\frac{Q}{V}t)}}
\end{equation}

The ventilation airflow rate was gathered from building HVAC data, and the temperature, concentration, and initial concentration were taken from the Awair sensor.The data used in the analysis consisted of the 15-minute period following the crushing of the plant leaves or the introduction of terpene essences into the room.

The emission rate was calculated for each chemical and dosage pair at each sensor. As the sensors are prone to error, the median VOC reading of the first 5 values was used for the initial concentration in the room. The emission rate was calculated for the last 5 values and the median of these values was used as the emission rate. The emission rates of the same chemical dosage pair were averaged for each sensor. 

Capturing all of the variables that help model the spread of VOCs in the room is difficult. Due to this, we use machine learning algorithms to determine the different terpenes and terpenoids in the room.

\subsubsection{Terpene Classification}
All Awair data were used from the dataset described in Section \ref{sec.data_collection}.
The dataset has control, 4 different unique chemicals, and a basil plant. We split the data into various groups for the machine learning model to classify. First, we observe all of the chemicals together vs. control (no terpene in the room) to see if machine learning can be used to determine the presence of a chemical in the room. Next, we observe the individual chemicals vs. control to see if the individual chemicals are distinguishable from `control'. 
Then, we look at the difference between 100 $\mu L$ and 200 $\mu L$ dosages of D-Limonene. Following the individual chemicals, we check if the machine learning models can distinguish between each of the chemicals. 
Finally, we verify if machine learning can be used to determine the plant stress tests from `control'. For each of the different tests, 20\% of the data was used in the testing set while the other 80\% was used as the training set. These data points are not enough to create a fully-fledged classification system, however, it can serve as a proof of concept for future work. 

Our classification-focused machine learning model employs the Random Forest, Support Vector Machine (SVM), and XGBoost algorithms, each chosen for its efficacy in handling classification tasks~\cite{schapire2015machine}. We chose to use all three of these to determine which worked better. Sklearn's Random Forest and SVM~\cite{scikitstable} were used and XGBoost's XGBClassifer~\cite{xgboost} was used.

\subsubsection{Feature Selection}

To extract features from the time series data, we used the {\textit{tsfresh}}~\cite{tsfresh} library, which is a Python library used to extract a large number of features from time series data. 
The list of all features can be found on their website~\cite{featureextract}.
As this library extracts numerous features, only the relevant features were selected using a function in tsfresh. 
Even with this functionality, there can be up to 300-400 relevant features. 
We select the best 15 features from these 300-400 by doing feature selection using SciKit's built-in \textit{SelectFromModel} class to prevent overfitting. We further analyzed the best features by employing Random Forest, SVM, and XGBoost, setting the stage for a comprehensive examination of their performance in subsequent tests.

After running all of the tests, we observe the best features for classification. 
As best 15 features are selected by the program for each test, they are not always the same. 
Although they are different, three show up in at least 7 out of the 9 machine learning tests. 
These features are the autocorrelation lag, permutation entropy, and approximate entropy. 
Autocorrelation lag finds the correlation between values that are a certain timestamp apart. 
All models use lag values between one and seven (10 seconds to 70 seconds). Permutation entropy captures the complexity of the system by ordering relations between values of time series data. 
Approximate entropy measures how unpredictable the fluctuations in the data set are. 
These features look at multiple sections of the time series data. These are different features than the normalized sensor response \cite{gancarz2019identification}, reaction time \cite{gancarz2019identification}, and median \cite{kocc2011role} suggested by previous studies.

\begin{table}[htb]
\caption{Table of all 16 Terpene Essences being tested in Experiment 1 for finding the highest $\Delta $ - Results concluded from four awair omni sensors in the room}
\label{tab.all_chemicals}
\resizebox{\columnwidth}{!}{
    \begin{tabular}{|c||c|c|c|}
    \hline
        Chemical &  Formula &  $\Delta_{final-initial}$~(ppb) & $\Delta_{max-initial}$~(ppb) \\
            \hline \hline
            $\alpha$-Bisabolol & $C_{15}H_{26}O$ & 235 & 257\\ 
            $\alpha$-Caryophyllene & $C_{15}H_{24}$ & 2 & 36\\ 
            $\alpha$-Phellandrene & $C_{10}H_{16}$ & 530 & 857\\ 
            $\alpha$-Pinene & $C_{10}H_{16}$ & 59 & 59\\ 
            $\alpha$-Terpinine & $C_{10}H_{16}$ & 1266 & 1811\\ 
            $\beta$-Caryophyllene & $C_{15}H_{24}$ & 42 & 44\\ 
            $\beta$-Pinene & $C_{10}H_{16}$ & 15 & 29\\ 
            Cedrene & $C_{15}H_{24}$ & 1 & 16\\ 
            Cis Beta-Ocimene 70\% & $C_{10}H_{16}$ & 38774 & 38774 \\ 
            Citral & $C_{10}H_{16}O$ & 1106 & 1106\\ 
            Citronellol & $C_{10}H_{20}O$ & 112 & 119\\ 
            D-Limonene & $C_{10}H_{16}$ & 4356 & 4356\\ 
            Delta-3-Carene  & $C_{10}H_{16}$ & 142 & 142\\ 
            Farnesene & $C_{15}H_{24}$ & 45 & 54\\ 
            Geranoil, Natural & $C_{10}H_{18}O$ & 45 & 290\\ 
            Linalool & $C_{10}H_{18}O$ & 23 & 27\\ 
            \hline
    \end{tabular}
    }
\end{table}

\section{Results}

In this section, we quantitatively answer the identified research questions through the three experiments. 
In section 3.1 we provided results from Experiment 1 and 2 and how we defined the terpene essences to test. 
In section 3.2 we delve more into the classification of the data through the physics and machine learning models. 
Lastly, in section 3.3 we present the results of experiment 3, where we examine VOC traces when plants are under normal and stressed conditions.

  

\begin{figure*}[!ht]
\vspace{-10pt}
\centering
\includegraphics[width=1\linewidth]{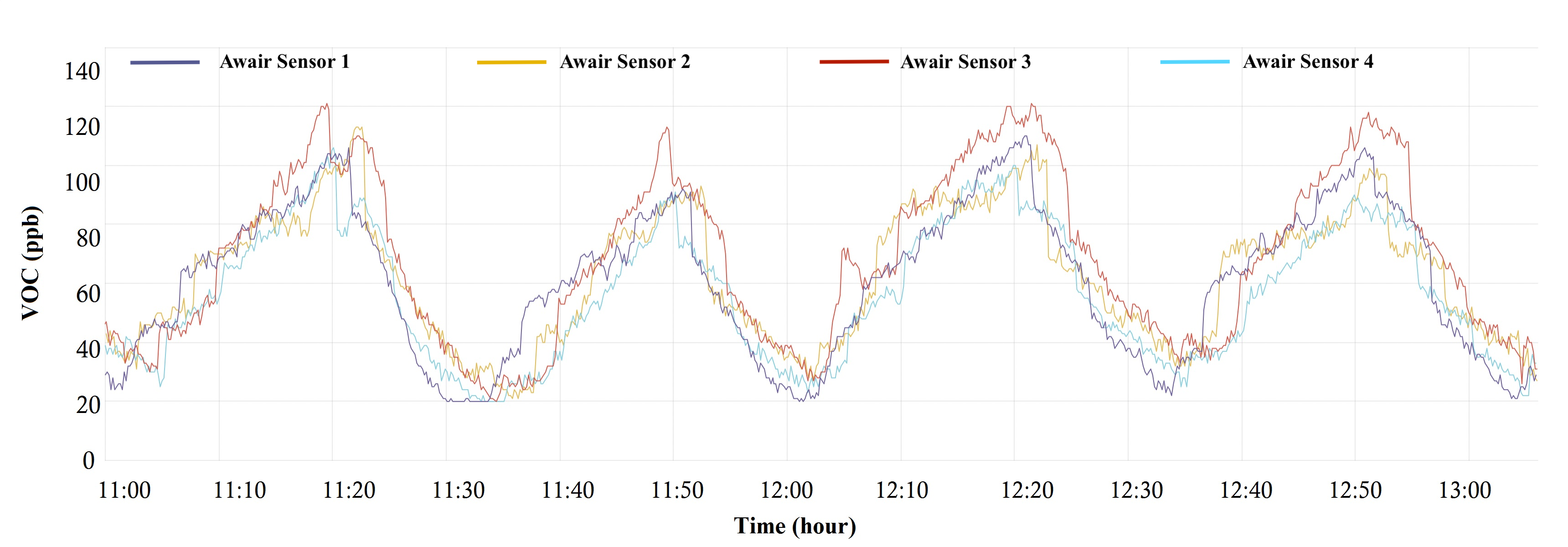}
\caption{Data collection by four sensors in Experiment 1 detecting D-limonene terpene signals across four consecutive experiments. Proximity to the terpene source correlates with higher maximum VOC trace points. Sensor 3 (the red line), positioned at a height of 75 cm, records the highest altitude among the sensors. Also, Awair Sensors 1 (purple line) and 4 (blue line) had the lowest picks}
\label{fig:Room_test_DL4}
\includegraphics[width=1\linewidth]{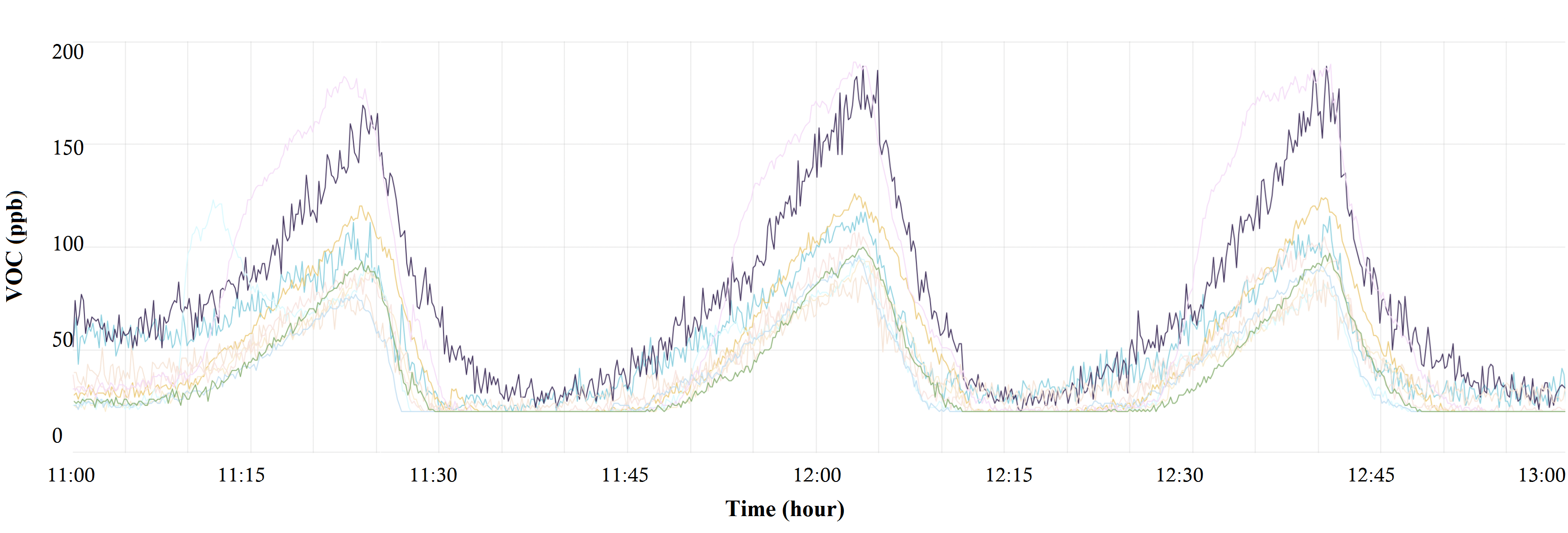}
\caption{Data collection by 13 sensors in Experiment 2 detecting D-limonene terpene signals across three consecutive experiments}
\label{fig:Room_test_DL13}
\vspace{-10pt}
\end{figure*}

\subsection{Terpene Detection}
To understand if the selected commercial sensor can detect terpenes we
performed Experiment 1 described in \cref{sec:method:base:awair}.
Table~\ref{tab.all_chemicals} shows the measured change in VOC readings
after introducing each terpene.
Of the 16 terpenes, Cis-Beta Ocimene had the best sensitivity and longevity with a
$\Delta_{max-initial}$ and $\Delta_{final-initial}$ of 38774~ppb. Following this were D-Limonene, Alpha-Terpinene, and Citral with $\Delta_{max-initial}$ and $\Delta_{final-initial}$ over 1000~ppb.

Overall, the terpenes with the best longevity and sensitivity were Cis-Beta Ocimene, D-Limonene, Citral, and Alpha-Terpinene.
These four terpenes share a similar chemical makeup (although Citral has an extra oxygen atom) and are all classified as monoterpenes. The Awair sensor detected all terpenes but with differing sensitivity levels. These results match the Awair website, as we were able to detect both D-Limonene and $\alpha$-pinene.
Also, due to the results obtained from the terpene essences tests conducted in Experiment 1 and the sensors' high sensitivity to Cis-Beta Ocimene and D-limonene, we opted to use a basil plant for Experiments 1 and 2. It is known that basil plants emit these two terpenes. 

For our complete dataset, we ran the experiment 81 times with 4 sensors (Experiment 1) and 85 times with 13 sensors ~(Experiment 2), with each trial to be about 15 minutes long. As a result, a total of 1429 (including 324 in Experiment 1 and 1105 in Experiment 2) VOC data points were in aggregate ~(one trace for each sensor).
This includes
481 Citral traces,
677 D-Limonene traces,
53 Cis-Beta Ocimene 70\% traces,
53 Alpha-Terpinene traces,
41 basil plant traces,
and 124 control traces.
In Experiment 1, for Citral, 12 traces used 100 $\mu L$, and 40 used 200 $\mu L$ dosage.
For D-Limonene, 40 traces used 100 $\mu L$ and 52 used 200 $\mu L$ dosage.
For Cis-Beta Ocimene 70\%, 12 traces used 100 $\mu L$ and 28 used 200 $\mu L$.
All Alpha-Terpinene traces 200 $\mu L$ dosage. 

On the other hand, in Experiment 2, all tests were done using 200 $\mu L$ dosage. 
A summary of all traces is in
Table~\ref{tab.dataset_summary}. 

Figures~\ref{fig:Room_test_DL4} and \ref{fig:Room_test_DL13} illustrate the results of D-limonene testing at 200 $\mu L$, showcasing the data collected from four and thirteen different air quality sensors in the room.

\begin{table}[htb]
        \caption{Summary of the Number of Experiments Conducted for Each Terpene Compound during Experiments 1 and 2}
        \label{tab.dataset_summary}
        \resizebox{\columnwidth}{!}{
            \begin{tabular}{|c||c|c|c|}
                \hline
                Terpene &  Total Traces & $200 \mu L$ Traces & $100 \mu L$ Traces \\ 
                    \hline \hline
                    Alpha-Terpinene & 53& 53& 0 \\
                    Cis-Beta Ocimene & 53& 41& 12  \\ 
                    Citral & 481& 469& 12 \\ 
                    D-Limonene & 677& 637& 40  \\ 
                    Control & 124& N/A & N/A  \\
                    Basil Plant& 41& N/A & N/A  \\
                    \hline
            \end{tabular}
            }
\end{table}

\begin{figure}[t]
    \centering
    \includegraphics[width=0.48\textwidth]{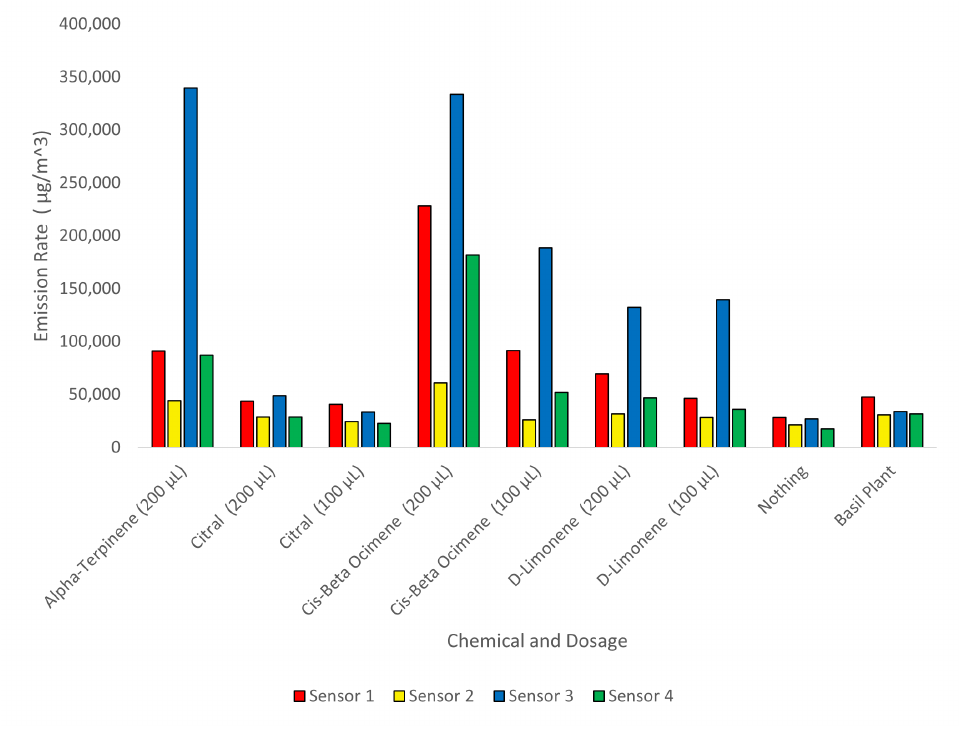}
    \caption{Emission rate graph for various terpenes and dosages detected by four sensors using the simple mass-based physics model. Sensor 3 consistently recorded the highest emissions, especially for 200~µL of Alpha-Terpinene, Citral, Cis-Beta Ocimene, and D-Limonene, with peaks exceeding 300,000~µg/m³. Sensors 1 and 2 reported significantly lower values, and controls (“Nothing” and “Basil Plant”) showed minimal emissions across all sensors (<50,000 µg/m³). The results reveal inconsistency in emission rates across sensors, deviating from the expected uniform readings. This inconsistency highlights the model's limitation, which fails to account for distance and sensor sensitivity. Consequently, the model is unsuitable for identifying terpenes in the room, suggesting that alternative equations may provide more accurate results.}
    \label{fig.emission_rate_bar_chart}
    \vspace{-15pt}
\end{figure}

\begin{table*}[htb!]
\caption{Machine Learning Classifier Results for Experiment 1 Dataset: Comparing the Accuracy of Different ML Models in Distinguishing Between Terpene Test Data Sets}

{Acc: Accuracy, Pre: Precision, Rec: Recall, F1: F1 Score}  
\label{tab.ml_results}
\begin{adjustbox}{width=\textwidth}
\begin{tabular}{|c|cccc|cccc|cccc|c|}
\hline
\multirow{2}{*}{Test}       & \multicolumn{4}{c|}{SVM}                                                                                & \multicolumn{4}{c|}{Random Forest}                                                                      & \multicolumn{4}{c|}{XGBoost}                                                                            & \multicolumn{1}{l|}{All Models} \\ \cline{2-14} 
                            & \multicolumn{1}{c}{Acc} & \multicolumn{1}{c}{Pre} & \multicolumn{1}{c}{Rec} & F-1 & \multicolumn{1}{c}{Acc} & \multicolumn{1}{c}{Pre} & \multicolumn{1}{c}{Rec} & F-1& \multicolumn{1}{c}{Acc} & \multicolumn{1}{c}{Pre} & \multicolumn{1}{c}{Rec} & F-1& Support\\ \hline
All Chemicals vs Control    & \multicolumn{1}{c }{93\%}      & \multicolumn{1}{c }{0.93}     & \multicolumn{1}{c }{0.93}   & 0.93      & \multicolumn{1}{c }{93\%}      & \multicolumn{1}{c }{0.95}     & \multicolumn{1}{c }{0.93}   & 0.94      & \multicolumn{1}{c }{93\%}      & \multicolumn{1}{c }{0.93}     & \multicolumn{1}{c }{0.93}   & 0.93      & 60                              \\ 
Cis-Beta Ocimene vs Control & \multicolumn{1}{c }{100\%}     & \multicolumn{1}{c }{1.00}     & \multicolumn{1}{c }{1.00}   & 1.00      & \multicolumn{1}{c }{100\%}     & \multicolumn{1}{c }{1.00}     & \multicolumn{1}{c }{1.00}   & 1.00      & \multicolumn{1}{c }{100\%}     & \multicolumn{1}{c }{1.00}     & \multicolumn{1}{c }{1.00}   & 1.00      & 23                              \\ 
D-Limonene vs Control       & \multicolumn{1}{c }{94\%}      & \multicolumn{1}{c }{0.94}     & \multicolumn{1}{c }{0.94}   & 0.94      & \multicolumn{1}{c }{97\%}      & \multicolumn{1}{c }{0.97}     & \multicolumn{1}{c }{0.97}   & 0.97      & \multicolumn{1}{c }{97\%}      & \multicolumn{1}{c }{0.97}     & \multicolumn{1}{c }{0.97}   & 0.97      & 33                              \\ 
Citral vs Control           & \multicolumn{1}{c }{96\%}      & \multicolumn{1}{c }{0.96}     & \multicolumn{1}{c }{0.96}   & 0.96      & \multicolumn{1}{c }{80\%}      & \multicolumn{1}{c }{0.80}     & \multicolumn{1}{c }{0.80}   & 0.80      & \multicolumn{1}{c }{80\%}      & \multicolumn{1}{c }{0.87}     & \multicolumn{1}{c }{0.80}   & 0.80      & 25                              \\ 
Alpha-Terpenine vs Control  & \multicolumn{1}{c }{100\%}     & \multicolumn{1}{c }{1.00}     & \multicolumn{1}{c }{1.00}   & 1.00      & \multicolumn{1}{c }{100\%}     & \multicolumn{1}{c }{1.00}     & \multicolumn{1}{c }{1.00}   & 1.00      & \multicolumn{1}{c }{100\%}     & \multicolumn{1}{c }{1.00}     & \multicolumn{1}{c }{1.00}   & 1.00      & 23                              \\ 
D-Limonene Dosages          & \multicolumn{1}{c }{37\%}      & \multicolumn{1}{c }{0.37}     & \multicolumn{1}{c }{0.37}   & 0.37      & \multicolumn{1}{c }{42\%}      & \multicolumn{1}{c }{0.43}     & \multicolumn{1}{c }{0.42}   & 0.42      & \multicolumn{1}{c }{42\%}      & \multicolumn{1}{c }{0.43}     & \multicolumn{1}{c }{0.42}   & 0.42      & 19                              \\ 
All Chemicals               & \multicolumn{1}{c }{62\%}      & \multicolumn{1}{c }{0.62}     & \multicolumn{1}{c }{0.62}   & 0.62      & \multicolumn{1}{c }{60\%}      & \multicolumn{1}{c }{0.61}     & \multicolumn{1}{c }{0.60}   & 0.59      & \multicolumn{1}{c }{60\%}      & \multicolumn{1}{c }{0.61}     & \multicolumn{1}{c }{0.60}   & 0.59      & 45                              \\ 
D-Lim vs Cis-Beta vs A-Terp & \multicolumn{1}{c }{66\%}      & \multicolumn{1}{c }{0.72}     & \multicolumn{1}{c }{0.66}   & 0.65      & \multicolumn{1}{c }{69\%}      & \multicolumn{1}{c }{0.68}     & \multicolumn{1}{c }{0.69}   & 0.67      & \multicolumn{1}{c }{63\%}      & \multicolumn{1}{c }{0.65}     & \multicolumn{1}{c }{0.63}   & 0.63      & 35                              \\ 
D-Lim vs Cis-Beta vs Citral & \multicolumn{1}{c }{43\%}      & \multicolumn{1}{c }{0.54}     & \multicolumn{1}{c }{0.43}   & 0.47      & \multicolumn{1}{c }{73\%}      & \multicolumn{1}{c }{0.73}     & \multicolumn{1}{c }{0.73}   & 0.73      & \multicolumn{1}{c }{70\%}      & \multicolumn{1}{c }{0.70}     & \multicolumn{1}{c }{0.70}   & 0.70      & 37                              \\ 
D-Lim vs Cis-Beta Ocimene   & \multicolumn{1}{c }{78\%}      & \multicolumn{1}{c }{0.83}     & \multicolumn{1}{c}{0.78}   & 0.74      & \multicolumn{1}{c}{89\%}      & \multicolumn{1}{c}{0.89}     & \multicolumn{1}{c }{0.89}   & 0.89      & \multicolumn{1}{c }{78\%}      & \multicolumn{1}{c }{0.77}     & \multicolumn{1}{c }{0.78}   & 0.77      & 27                              \\ 
Plants vs Control           & \multicolumn{1}{c}{75\%}      & \multicolumn{1}{c }{0.80}     & \multicolumn{1}{c}{0.75}   & 0.76      & \multicolumn{1}{c}{90\%}      & \multicolumn{1}{c}{0.90}     & \multicolumn{1}{c}{0.90}   & 0.90      & \multicolumn{1}{c}{95\%}      & \multicolumn{1}{c}{0.96}     & \multicolumn{1}{c }{0.95}   & 0.95      & 20                              \\ \hline

\end{tabular}%
\end{adjustbox}
\end{table*}

\begin{table*}[htb!]
\caption{Machine Learning Classifier Results for Experiment 2 Dataset: Comparing the Accuracy of Different ML Models in Distinguishing Between Terpene Test Data Sets}

\label{tab.ml_results13}
\begin{adjustbox}{width=\textwidth}
\begin{tabular}{|c|cccc|cccc|cccc|c|}
\hline
\multirow{2}{*}{Test}       & \multicolumn{4}{c|}{SVM}                                                                                & \multicolumn{4}{c|}{Random Forest}                                                                      & \multicolumn{4}{c|}{XGBoost}                                                                            & \multicolumn{1}{l|}{All Models} \\ \cline{2-14} 
                            & \multicolumn{1}{c}{Acc} & \multicolumn{1}{c}{Pre} & \multicolumn{1}{c}{Rec} & F-1& \multicolumn{1}{c}{Acc} & \multicolumn{1}{c}{Pre} & \multicolumn{1}{c}{Rec} & F-1 & \multicolumn{1}{c}{Acc} & \multicolumn{1}{c}{Pre} & \multicolumn{1}{c}{Rec} & F-1 & Support\\ \hline
All Chemicals vs Control    & \multicolumn{1}{c}{96\%}      & \multicolumn{1}{c}{0.96}     & \multicolumn{1}{c}{0.96}   & 0.96& \multicolumn{1}{c}{99\%}      & \multicolumn{1}{c}{0.99}     & \multicolumn{1}{c}{0.99}   & 0.98& \multicolumn{1}{c}{97\%}      & \multicolumn{1}{c}{0.97}     & \multicolumn{1}{c}{0.97}   & 0.96& 211\\ 
Cis-Beta Ocimene vs Control & \multicolumn{1}{c}{100\%}     & \multicolumn{1}{c}{1.00}     & \multicolumn{1}{c}{1.00}   & 1.00      & \multicolumn{1}{c}{100\%}     & \multicolumn{1}{c}{1.00}     & \multicolumn{1}{c}{1.00}   & 1.00      & \multicolumn{1}{c}{85\%}     & \multicolumn{1}{c}{0.91}     & \multicolumn{1}{c}{0.85}   & 0.86& 13\\ 
D-Limonene vs Control       & \multicolumn{1}{c}{90\%}      & \multicolumn{1}{c}{0.88}     & \multicolumn{1}{c}{0.90}   & 0.88& \multicolumn{1}{c}{91\%}      & \multicolumn{1}{c}{0.92}     & \multicolumn{1}{c}{0.91}   & 0.89& \multicolumn{1}{c}{91\%}      & \multicolumn{1}{c}{0.90}     & \multicolumn{1}{c}{0.91}   & 0.90& 128\\ 
Citral vs Control           & \multicolumn{1}{c}{91\%}      & \multicolumn{1}{c}{0.91}     & \multicolumn{1}{c}{0.91}   & 0.91& \multicolumn{1}{c}{97\%}      & \multicolumn{1}{c}{0.97}     & \multicolumn{1}{c}{0.97}   & 0.96& \multicolumn{1}{c}{98\%}      & \multicolumn{1}{c}{0.98}     & \multicolumn{1}{c}{0.98}   & 0.98& 94\\ 
Alpha-Terpenine vs Control  & \multicolumn{1}{c}{100\%}     & \multicolumn{1}{c}{1.0}     & \multicolumn{1}{c}{1.00}   & 1.00      & \multicolumn{1}{c}{85\%}     & \multicolumn{1}{c}{0.91}     & \multicolumn{1}{c}{0.85}   & 0.86& \multicolumn{1}{c}{100\%}     & \multicolumn{1}{c}{1.0}     & \multicolumn{1}{c}{1.0}   & 1.0& 13\\ 
D-Limonene  vs Citral& \multicolumn{1}{c}{70\%}      & \multicolumn{1}{c}{0.70}     & \multicolumn{1}{c}{0.70}   & 0.70& \multicolumn{1}{c}{86\%}      & \multicolumn{1}{c}{0.86}     & \multicolumn{1}{c}{0.86}   & 0.86& \multicolumn{1}{c}{77\%}      & \multicolumn{1}{c}{0.77}     & \multicolumn{1}{c}{0.77}   & 0.76& 201\\ 
All Chemicals               & \multicolumn{1}{c}{67\%}      & \multicolumn{1}{c}{0.67}     & \multicolumn{1}{c}{0.67}   & 0.67& \multicolumn{1}{c}{86\%}      & \multicolumn{1}{c}{0.86}     & \multicolumn{1}{c}{0.86}   & 0.86& \multicolumn{1}{c}{67\%}      & \multicolumn{1}{c}{0.67}     & \multicolumn{1}{c}{0.67}   & 0.67& 206\\ 
D-Lim vs Cis-Beta vs A-Terp & \multicolumn{1}{c}{97\%}      & \multicolumn{1}{c}{0.72}     & \multicolumn{1}{c}{0.66}   & 0.65      & \multicolumn{1}{c}{96\%}      & \multicolumn{1}{c}{0.96}     & \multicolumn{1}{c}{0.96}   & 0.95& \multicolumn{1}{c}{96\%}      & \multicolumn{1}{c}{0.95}     & \multicolumn{1}{c}{0.96}   & 0.95& 123\\ 
D-Lim vs Cis-Beta vs Citral & \multicolumn{1}{c}{70\%}      & \multicolumn{1}{c}{0.70}     & \multicolumn{1}{c}{0.70}   & 0.70& \multicolumn{1}{c}{86\%}      & \multicolumn{1}{c}{0.86}     & \multicolumn{1}{c}{0.86}   & 0.85& \multicolumn{1}{c}{79\%}      & \multicolumn{1}{c}{0.78}     & \multicolumn{1}{c}{0.79}   & 0.78& 203\\ 
D-Lim vs Cis-Beta Ocimene   & \multicolumn{1}{c}{97\%}      & \multicolumn{1}{c}{0.97}     & \multicolumn{1}{c}{0.97}   & 0.96& \multicolumn{1}{c}{97\%}      & \multicolumn{1}{c}{0.97}     & \multicolumn{1}{c}{0.97}   & 0.97& \multicolumn{1}{c}{96\%}      & \multicolumn{1}{c}{0.95}     & \multicolumn{1}{c}{0.96}   & 0.95& 123\\ 
Plants vs Control           & \multicolumn{1}{c}{100\%}      & \multicolumn{1}{c}{1.0}     & \multicolumn{1}{c}{1.0}   & 1.0& \multicolumn{1}{c}{100\%}      & \multicolumn{1}{c}{1.0}     & \multicolumn{1}{c}{1.0}   & 1.0& \multicolumn{1}{c}{100\%}      & \multicolumn{1}{c}{1.0}     & \multicolumn{1}{c}{1.0}   & 1.0& 13\\ \hline

\end{tabular}%
\end{adjustbox}
\end{table*}

    

\begin{figure*}[h!]
    \vspace{-10pt}
    \centering
    \begin{subfigure}[b]{0.8\textwidth}
        \centering
        \includegraphics[width=\textwidth]{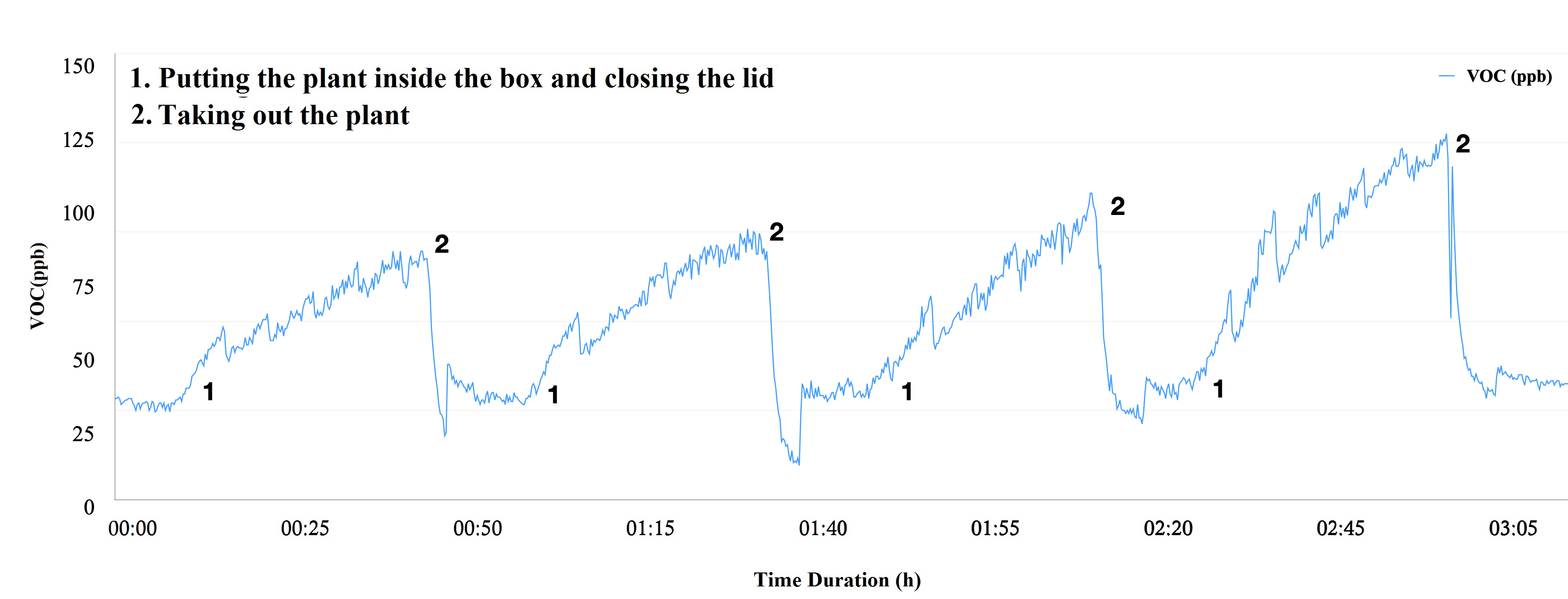}
        \caption{VOC response when the plant is not stressed}
        \label{fig:box_tests1a}
    \end{subfigure}
    \vspace{0.5em} 
    
    \begin{subfigure}[b]{0.8\textwidth}
        \centering
        \includegraphics[width=\textwidth]{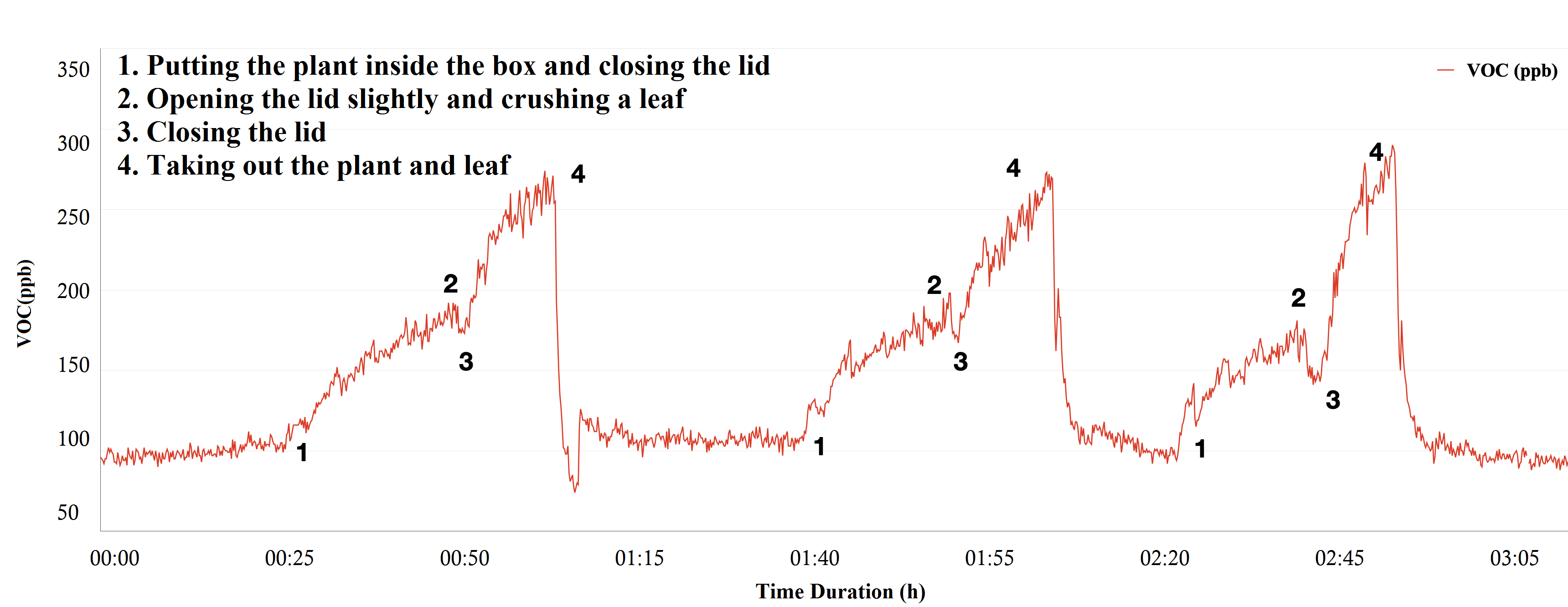}
        \caption{VOC response when the plant is stressed}
        \label{fig:box_tests1b}
    \end{subfigure}
    
    \caption{VOC responses of the plant under no stress and stress. Each specific experiment was performed 100 times. \cref{fig:box_tests1a} shows the response when the plant is placed in the control box and is not subject to any stress. Each time the plant is placed in the box, there is an increase in the VOC levels, but it stabilizes before being removed. The peak VOC readings are around 225ppm, but there is an accumulation of the VOC concentration the longer the plant is placed in the box. \cref{fig:box_tests1b} shows the response when the plant is placed in the control box and is subjected to stress by crushing its leaves. This plot shows the response of the plant when one of its leaves is crushed, the VOC response shows a small dip~(stages 2 to 3 marked in the graph) because the control box is opened, resulting in VOCs escaping outside, however, once the box is closed, the VOC response shows a sharp spike, in response to the stress. This experiment is performed 3 times. This contrast between~(a) and~(b) showcases that the sensor is able to pick up on the VOC emission when the plant is stressed.}
    \label{fig:box_tests1}
    \vspace{-15pt}
\end{figure*}

\begin{figure*}[t!]
    \vspace{-10pt}
    \centering
    \begin{subfigure}[b]{0.9\textwidth}
        \centering
        \includegraphics[width=\textwidth]{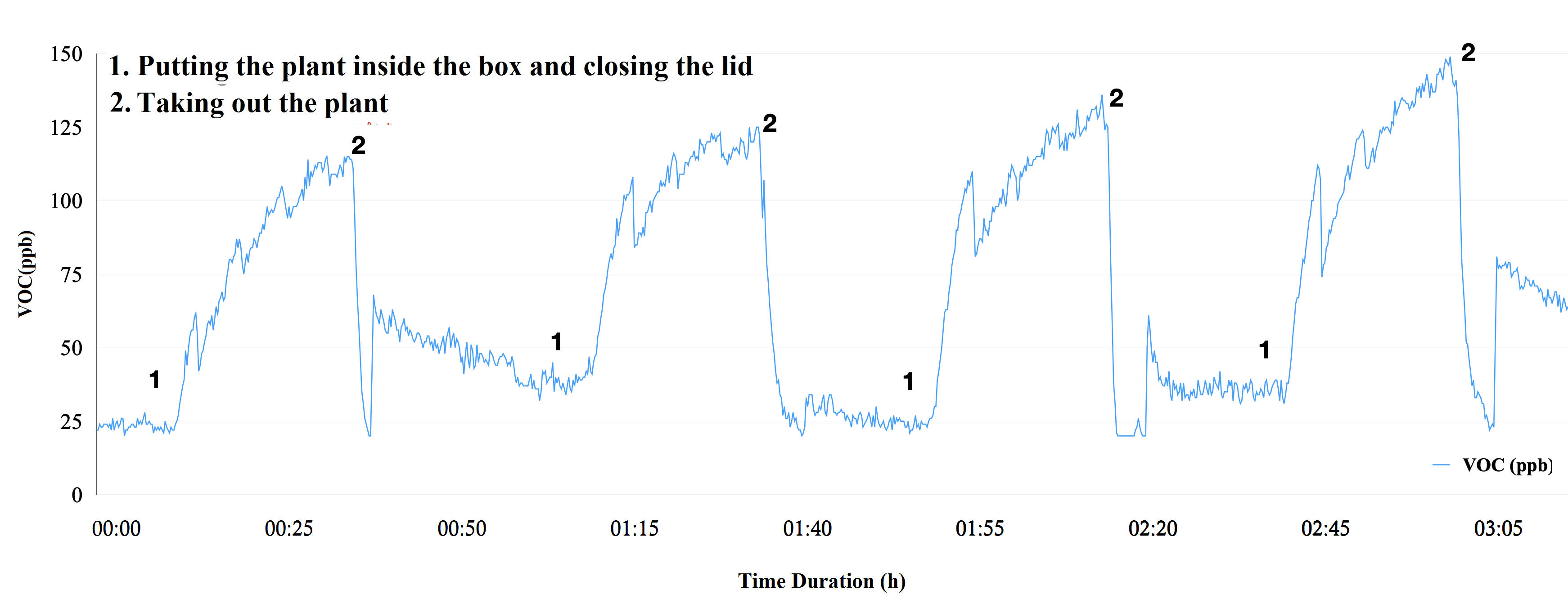}
        \caption{VOC response when the plant is not stressed}
        \label{fig:box_testsa}
    \end{subfigure}
    \vspace{0.5em} 
    
    \begin{subfigure}[b]{0.9\textwidth}
        \centering
        \includegraphics[width=\textwidth]{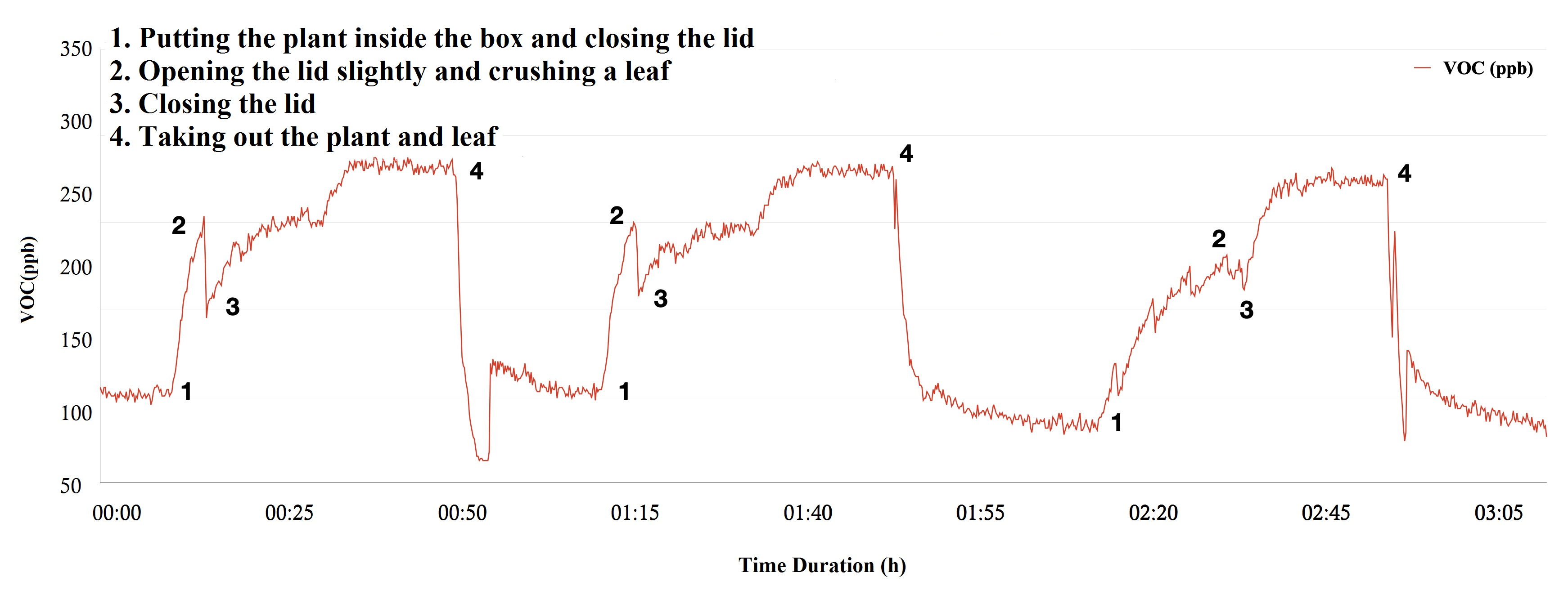}
        \caption{VOC response when the plant is stressed}
        \label{fig:box_testsb}
    \end{subfigure}
    
    \caption{The same experiment from \cref{fig:box_tests1} performed on a different day, showcases the same results. The values of the VOC readings are different from the previous figure, but this is because of the temperature and humidity difference between the two days. Despite this, the VOC profiles are similar and showcase the same behavior. This leads us to conclusively determine that the commercial sensor is able to pick up on plant terpene signals.}
    \label{fig:box_tests}
    \vspace{-15pt}
\end{figure*}



\subsection{Classification Performance}

\subsubsection{Physics Model}

We calculate the average emission rates for each dosage and chemical pair at each sensor using the physics model from \cref{phymodel} and show them in \cref{fig.emission_rate_bar_chart}. In this Figure, the emission rates of four sensors in Experiment 1 were calculated. The emission rate at all sensors greatly varies (100,00 - 350,000 $\mu g$/$m^{3}$) for Alpha-Terpinene (200$\mu L$), Cis-Beta Ocimene 
(200$\mu L$ and 100$\mu L$) and D-Limonene (200$\mu L$ and 100$\mu L$). There was a slight variation (25,000 - 50,000 $\mu g$/$m^{3}$) in the emission rates at all sensors for Citral~(200$\mu L$ and 100$\mu L$), basil, and control. Also, all terpene essences and the basil plant had higher emission rates than the "control" in the room. We acknowledge that there are other physics models which may provide a better estimate of the VOC spread. Also, based on Figure 7, in most of the tests, the closest sensor (awair Sensor 3) had the highest reading among all other sensors, but we can't find a linear relationship between the sensor's readings and the distance from the VOC source.

\subsubsection{Machine Learning}

The accuracy results for the Random Forest, SVM, and XGBoost classifiers can be seen in Tables \ref{tab.ml_results} and \ref{tab.ml_results13}. Table \ref{tab.ml_results} is the results from Experiment 1 and Table \ref{tab.ml_results13} shows the results of Experiment 2. 


For the VOC-time data collected from  Experiment 1, the models successfully distinguished between the presence of any terpene in the room and the control, specifically Cis-Beta Ocimene and D-Limonene, as well as between the presence of plants and the control.  Among these models, Random Forest and XGBoost outperformed the SVM model, achieving the highest accuracies. Table 3 presents the results of machine learning classifier models applied to distinguish between different terpene test datasets in Experiment 1. Across various tests comparing terpene datasets with control groups or among different terpenes, SVM, Random Forest, and XGBoost classifiers achieved notable accuracies ranging from 37\% to 100\%. For instance, in distinguishing between Cis-Beta Ocimene and the control group, all three classifiers achieved a perfect accuracy of 100\%. Similarly, for Alpha-Terpenine versus the control group, the classifiers again achieved 100\% accuracy. However, in more nuanced tests such as D-Limonene Dosages, where different concentrations of D-Limonene were compared, the accuracies dropped to 37\%, 42\%, and 42\% for SVM, Random Forest, and XGBoost respectively, indicating the challenge of discerning subtle variations in terpene concentrations. Overall, these results demonstrate the efficacy of machine learning classifiers in accurately identifying and categorizing terpene datasets, with performance varying depending on the complexity of the comparison.

In Experiment 2, with an increased number of sensors (13) and a greater emphasis on D-limonene vs. Citral tests, the accuracy improved for all experiments except for D-limonene vs. Control. The disparity arises from the fact that there were over 10 times as many D-Limonene tests as control tests, whereas in Experiment 1, the numbers of each test were nearly equal and significantly lower than those in experiment 2. 
Additionally, we replaced the D-Limonene Dosages test with a new test comparing D-Limonene vs. Citral.

Based on the machine learning training results (Table  \ref{tab.ml_results13}) across different comparisons, SVM, Random Forest, and XGBoost classifiers achieved high accuracies ranging from 67\% to 100\%. 

we achieved an accuracy of 86\% in distinguishing between D-Limonene and Citral. Similarly, an accuracy of 86\% is achieved distinguishing between D-Limonene, Cis-Beta, and Citral. Notably, in tests such as Cis-Beta Ocimene versus the control group and Alpha-Terpenine versus the control group, all three classifiers consistently achieved perfect accuracies of 100\%, indicating distinct separability of these terpenes from the control datasets. However, in more nuanced differentiations like D-Limonene versus Citral, accuracies ranged from 70\% to 86\%, suggesting a relatively higher complexity in distinguishing between these specific terpenes. Moreover, when considering the ranking of the best accuracies, tests involving comparisons with the control group, particularly Cis-Beta Ocimene versus control and Alpha-Terpenine versus control, consistently demonstrated the highest accuracies (100\%) across all three classifiers, underscoring their distinctiveness. Furthermore, we were able to identify the differences between D-Limonene, Cis-Beta, and A-terp with an accuracy of 96\% and nearly 100\% accuracy when detecting the presence of plants in the room. Overall, these findings underscore the effectiveness of machine learning classifiers in accurately identifying and categorizing terpene datasets, with performance varying based on the complexity of the comparison and the distinctiveness of the terpenes involved.

After running all of the tests, we consider the best features for classification. As the best 15 features are selected by the algorithm for each test, they are not always the same. Although they are different, three features show up in at least 7 out of the 9 machine learning tests. These features are the autocorrelation lag, permutation entropy, and approximate entropy. 
Autocorrelation lag finds the correlation between values that are a certain timestamp apart. All models use lag values between 1 and 7~(10 seconds to 70 seconds). Permutation entropy captures the complexity of the system by capturing order relations between values of time series data. Approximate entropy measures how unpredictable the fluctuations in the data are. All of these features look at multiple sections of the time series data. These are different features than the normalized sensor response, reaction time, and median suggested by previous studies~\cite{gancarz2019identification, kocc2011role}.

\subsection{Plant Terpene Trace Interpretation}

After detecting Awair sensor responses and classifying terpenes, including terpene essences and basil plant terpenes, in Experiments 1 and 2, we observed repeating patterns across consecutive tests. 
We designed Experiment 3 to closely examine the VOC-time traces of terpene emissions. For this purpose, we established a controlled environment within a confined space, namely a box, to investigate the terpene emission traces of a live basil plant, its interpretation, and differentiation in normal vs stressed conditions. 


We utilized an average-sized basil plant with approximately 60 leaves, positioning it inside the testing Box. \cref{fig:box_tests1a} displays the emission rates of the plant when undisturbed, while \cref{fig:box_tests1b} illustrates the emission rate of the plant under stress, achieved by crushing a leaf.
The baseline of the testing environment can change from 20 ppb to 120 ppb based on the factors we discussed in the methodology. We replicated the experiment on separate days with different VOC baselines. For example, the baseline for basil plant in normal condition for \cref{fig:box_tests1a} and \cref{fig:box_testsa} were 30 ppb and 25 ppb, respectively for two different days. Also, for a stressed basil plant while crushing a leaf, the baseline of the testing environment was 90 and 110 as shown in \cref{fig:box_tests1b} and \cref{fig:box_testsb}. 


In both conditions, with no leaves crushed and one leaf crushed, we observed a significant increase in VOC levels, which was 80-100 ppb for the basil plant with no stress and 150-180 ppb for the basil plant with stress. This observation suggests the accumulation of VOCs in the enclosed environment over time.
\cref{fig:box_tests1a} and \cref{fig:box_testsa} depict VOC levels when the plant was not subjected to stress during each 20-minute experiment window. Following the introduction of the basil plant, the sensors registered an increase of 100 to 150 ppb in VOC levels. 
This suggests that the plant can consistently produce approximately 100-150 ppb of VOCs within a 20-minute timeframe. It's important to note that this production may vary due to factors like sunlight exposure, temperature, humidity, and the time of day, all of which can influence the plant's VOC emissions

\cref{fig:box_tests1b} and \cref{fig:box_testsb} illustrate the effect of crushing a leaf on the VOC readings. Upon crushing a leaf, the VOC levels exhibit a noticeable increase of around 70 to 100 ppb, indicating that the plant releases a substantial quantity of VOCs, particularly terpenes, into the surrounding environment when subjected to leaf crushing as a stressor.
These findings not only demonstrate the ability of commercial sensors to detect and track changes in VOCs within the environment but also establish a link between plants and their terpene production.
By crushing a leaf, we directly influence the VOC profile within a confined environment, and this change can be detected and quantified using commercial sensors. In \cref{fig:box_tests1b} and \cref{fig:box_testsb}, steps numbers 2 and 3 can show how crushing a leaf can change the profile from 160 ppb to 260 ppb. While this test can show how a crushed leaf can influence plant emission, this discovery highlights the potential of utilizing simple, commercially available sensors in studying plant VOC emissions and their dynamics, offering valuable insights into plant-terpene interactions. In this study, our Awair sensor had a resolution of 1ppb and an accuracy of ±15\% which can be ignored because of the scale of VOC readings in our study which is 10 ppb.

The repetition of consecutive tests demonstrates the similarity in test patterns. These traces serve as a distinctive signature for each terpene during testing, and they may vary in different test environments. However, they consistently maintain a similar trace with varying frequencies and amplitudes. When discussing the VOC trace, there are several factors contributing to these traces that we should take into account when interpreting the Figures ~\ref{fig:Room_test_DL4},  ~\ref{fig:Room_test_DL13}, \ref{fig:box_tests1}, and \ref{fig:box_tests}. The factors that shape these patterns include the VOC baseline, VOC max point (amplitude of the signal), VOC rate (slope of the signal), and sensor placement in the room, based on the experiment scope of this study and their results, VOC baseline, VOC max point and Sensor placement in the room were discussed based in details in the subsequent sections:

\subsubsection{VOC baseline}
The primary factor contributing to the VOC baseline in a room is the quantity of equipment within that room producing VOCs. Our extensive testing has revealed that the baseline levels of VOCs in the testing environment, encompassing both the Room and the Box, can be impacted by various factors. In addition to the introduction of new VOC sources or alterations in human activities, several pivotal factors, including the number of equipment present, as well as temperature and humidity, significantly influence the definition of the VOC baseline in the testing environment. It is observed that a higher number of equipment in a room correlates with an elevated baseline of VOCs in that room. 

During our data analysis, we strive to select experiments in which the baseline conditions remain consistent. However, we acknowledge that temperature and humidity can also impact the analysis. 

The VOC baseline of the room we were using for Experiments 1 and 2 has the baseline range of 100 ppb to 200 ppb in different days, but mostly it was around 150 ppb. Also, the VOC baseline of the box we used for Experiment 3 had a baseline range from 10-40 ppb each time of the day we were doing tests in it. It is important to consider and remove the VOC baseline effect from the VOC readings.  


\subsubsection{VOC max point (amplitude of the signal)}
The maximum level of VOC that sensors can detect in each test depends on several factors:

\begin{enumerate}
    \item The proximity of the sensor to the VOC source.
    \item Temperature and humidity of the room.
    \item The baseline VOC levels in the room.
    \item The VOC type and dosage of the source
    \item The sensor age.
    \item The amount of time the VOC source is in the room 
    \item The sensitivity and selectivity of the VOC sensor.
    \item Air change rate of the room. 

\end{enumerate}
 In all experiments, the maximum point on the VOC diagram could vary from 100 to 10,000 ppb, depending on factors such as the type and amount of terpene present, as well as the exposure time to the room, with a daily VOC baseline ranging from 20 to 150 ppb. The difference between the max VOC point and baseline corresponds to the VOC dosages of 100 $\mu L$ and 200 $\mu L$ which can vary from 100 ppb to 200 ppb for 15 minutes of terepne exposure.
 Also, across the days that no experiment were conducted, the difference between the max VOC point on the and the baseline for each day remained relatively constant. 

Also, the sensors distance is important in collecting the VOC. For example, in the case of D-limonene, this VOC dosage could vary within the range of 50 to 155 ppb, depending on the sensor's distance within our testing room. 

Subsequently, in Experiment 3, we explore how crushing a leaf of the plant can release approximately the same amount of VOC which 120-150 ppb. This discovery guided our decision to just select  200 $\mu L$ as the volume dosage for testing VOC terpene quantities in Experiment 2.

\subsubsection{Slope of the VOC}

The slope of the VOC-time diagram is another critical factor in understanding the dynamics of VOC emissions. The slope of the VOC-time diagram reflects the rate of change in VOC concentration over time, which provides insights into key processes and events during the experiment. For example in Figure 9, sharp negative slopes, such as at 4 (removing the plant), highlight the removal or dilution of VOCs from the system. This occurs due to ventilation and the absence of a continuous VOC source after the plant is taken out. Also, steep positive slopes, such as during phase 2 (opening the lid and crushing a leaf), indicate a rapid release of VOCs into the environment. This suggests that physical interactions with the plant, like crushing a leaf, lead to significant and immediate emission of terpenes or other VOCs stored in plant tissues. The rate of this increase can be linked to the intensity of the disturbance and the plant’s VOC emission capacity. The slope of a VOC-time diagram is influenced by several factors, including the distance between the sensor and the source, the air exchange rate, the strength of the VOC emission, and the level of air mixing in the environment. A steeper slope indicates a faster rate of change in VOC concentration, which is typically observed when the sensor is placed closer to the source, the airflow is minimal, or the source releases VOCs rapidly (e.g., when crushing a leaf). Understanding these factors allows researchers to optimize sensor placement and environmental conditions to capture precise data. For instance, sharper slopes can help detect subtle emission patterns or rapid responses, aiding in studies of plant VOC release under different stress conditions or in assessing the efficiency of air filtration systems.

\subsubsection{Sensor placement in the room }

The illustration in Figure~\ref{fig:Room_test_DL4} and ~\ref{fig:Room_test_DL13} elucidates the significant influence of sensor placement on detecting VOC levels originating from the terpene source within the room. It's often assumed that the farther a sensor is from the source, the lower the observed VOC levels, but this oversimplification doesn't always hold true due to various factors influencing sensor readings as indicated in section 3.3.2. To gain a more comprehensive and quantitative insight into the significance of sensor placement, we conducted machine learning classification experiments for each of the 13 sensors in Experiment 2, with results presented in Table 5. Table ~\ref{tab.Sensor_summary}  presents accuracy results for RF, SVM, and XGBoost classifiers across 13 sensor locations, aiming to determine optimal sensor placement for terpene datasets. Notably, the sensor placed on the cabinet (C) exhibits the highest accuracy at 81\%, outperforming other locations such as the left wall up (LU) at 75\% and the return air (RA) at 69\%. Conversely, sensors located on the front wall next to the window (FW) and the supply air (SA) demonstrate relatively lower accuracies at 69\% and 56\% respectively.



\begin{table}[]
\tiny
\centering
\caption{Sensor Location summary: Accuracy results for the RF, SVM, and XGBoost classifiers for each of the 13 sensors individually, considering all terpene datasets to determine the optimal sensor placement. }
\label{tab.Sensor_summary}
\resizebox{\columnwidth}{!}{
    \begin{tabular}{|c||c|c|c|} \hline 
        
          \multicolumn{2}{|c||}{Sensor Location}& Symbol& ML Accuracy\\ \hline  \hline
            
             \multicolumn{2}{|c||}{Table 1}& T1& 0.62\\ 
             \multicolumn{2}{|c||}{Table 2}& T2& 0.38\\   
             \multicolumn{2}{|c||}{Table 3}& T3& 0.56\\  
             \multicolumn{2}{|c||}{Table 4}& T4& 0.69\\  
             \multicolumn{2}{|c||}{On the Cabinet}& C& 0.81\\ 
             \multicolumn{2}{|c||}{Front wall next to window}& FW& 0.69\\ 
                    
  \multicolumn{2}{|c||}{Right wall}& RW& 0.62\\ 
  \multicolumn{2}{|c||}{Left wall right side}& LR& 0.69\\ 
  \multicolumn{2}{|c||}{Left wall left side}& LL& 0.56\\
  \multicolumn{2}{|c||}{Left wall up}& LU& 0.75\\
  \multicolumn{2}{|c||}{On the Floor}& F& 0.62\\
  \multicolumn{2}{|c||}{Return Air (output)}& RA& 0.69\\
  \multicolumn{2}{|c||}{Supply Air (input)}& SA& 0.56\\\hline
            \end{tabular}
            }
            
\end{table}

Based on accuracy alone, optimal sensor placement tends to favor locations close to the sample. However, in buildings with multiple VOC sources, it is impractical to place a sensor right next to each source due to the sheer number of sources and the associated cost of acquiring numerous sensors. Therefore, relying solely on accuracy is insufficient in determining the best sensor location. In our context, the ideal sensor location achieves high accuracy in identifying different VOC sources while being positioned equidistant from all sources. Hence, both distance and accuracy are crucial factors in defining the optimal sensor placement. So, when both distance and accuracy are taken into account, the Return Air location emerges as the most suitable. Despite being the farthest from the sample, it exhibits the highest accuracy among all the other far-distance locations. Notably, sensors positioned near the sample, such as the cabinet and left upper sensor, demonstrate commendable accuracies at 0.81 and 0.75 respectively. Additionally, locations including Return Air, Left Wall Right Side, Front Wall Next to Window, and Table Number 4 sensors share a notable accuracy of 0.69. Yet, when considering their distances, Return Air stands out for maintaining similar accuracy despite its greater distance. Consequently, Return Air is identified as the optimal location among similarly ranked positions. Thus, considering the baseline accuracy of sensors close to and above the sample, Return Air with an accuracy of 0.69\% emerges as the most favorable sensor placement within the room. This conclusion is logical, given the numerous VOC sources in the environment, coupled with the impracticality of relocating sensors frequently. Thus, the "Return Air" location presents a compelling choice, where gases and particles converge for removal through the ventilation system.

\section{Discussion}

The first question of our study aimed to assess the efficacy of commercial sensors in detecting terpenes within indoor environments. Experiment 1 focused on evaluating the capability of Awair Omni sensors to detect terpenes. Through our experimentation with various terpene essences in an office setting, as seen from the results of Table 1, the commercial sensors was able to detect Cis-Beta Ocimene, D-Limonene, Alpha-Terpinene, and Citral the most and for the longest amount of time. Although they have the same chemical formulae~(other than Citral), the structure of the molecules is not the same. Different structures have an impact on the normalized sensor response and reaction time of a sensor~\cite{gancarz2019identification}. Furthermore, chemicals with different formulas have different sensor responses~\cite{ponzoni2008bread}. With regard to plants, basil plants are large emitters of Cis-Beta Ocimene and D-Limonene, making them useful plants for commercial sensors.


\vspace{2mm}
\textit{\textbf{Key Takeaway:
Sensor longevity and sensitivity are unique to each terpene.}} 
\vspace{2mm}

To answer question 2, Experiment 2 aimed to leverage physics-based and machine learning models to classify and differentiate terpenes based on sensor data. By expanding our investigation and strategically positioning sensors within the room, we sought to enhance the accuracy of our classification models. Our analysis revealed that Cis-Beta Ocimene and D-Limonene yielded the most promising results, making it the focal point of our study due to its abundance in basil. The live plant emission test results demonstrated that we can identify plant traces (signatures) based on the terpenes they release and the number of terpenes emitted when subjected to leaf crush stress, utilizing commercial gas sensors. Moving forward, the live emission test involving various plants will serve as a platform for future research, enabling us to investigate the terpenes released by different plants, explore different stressors affecting plants, and examine their corresponding responses.


\vspace{2mm}
\textit{\textbf{Key Takeaway: Terpenes do not have the same emission rate at different sensors when using our simple mass-based equation.}} 
\vspace{2mm}

Overall, the use of machine learning allowed us to accurately determine if there is any chemical in the room, a specific chemical in the room, which chemical is in the room between 2 chemicals, and if a plant is in the room. The tests against control in the room which is the control versus something in the room that can explain the change in the VOC readings of the room. Naturally, the composition of VOCs in a room changes naturally throughout the day even when there is nothing present in it~\cite{hewitt2022variation}. Although there is a change, it is not as drastic as when a terpene is in the room. This makes it easier to identify when using features such as the autocorrelation lag, permutation entropy, and approximate entropy. With regard to determining which of the two chemicals is in the room, expensive sensors can detect unique responses to different VOCs~\cite{dragonieri2017electronic}. This demonstrates that sensors can have unique responses to the chemicals, even with the lack of sensitivity provided by lower-costing sensors~\cite{spinelle2017review}. If the responses are unique enough and there are limited chemicals to choose from, the machine learning model has the ability to determine the difference.

In Experiment 1, we faced challenges in identifying the specific chemical present in the room when choosing between 3 or 4 chemicals, and we struggled to accurately determine the dosage of the chemical released. However, during the 13-sensor experiment 2, we significantly improved our ability to distinguish between different chemicals in the room across all comparison tests. Due to the lack of sensitivity of lower cost sensors~\cite{spinelle2017review}, commercial sensors are not as accurate as more expensive E-nose sensors, especially when not in a testbed designed to get the best VOC readings. E-nose sensors are able to distinguish the differences between different volatile profiles~\cite{delgado2012use}. The lack of an expensive sensor combined with the nature of VOC curves to be different with consistent dosages and times~\cite{yeoman2021estimating} can make it challenging to collect similar enough data each time. This can cause accuracy to go down when using terpenes with similar "ideal" VOC curves, as the sensor insensitivity and environment can cause variations in the readings.

\vspace{2mm}
\textit{\textbf{Key Takeaway:
Machine learning can be used with commercial sensors to identify some individual terpenes.\\
The best features for identification with commercial sensors look at sections of the VOC curve.}} 
\vspace{2mm}

Ultimately, our utilization of machine learning to assess the accuracy of each sensor relative to the best sensor underscores the substantial impact of sensor placement on the detection of VOC levels emanating from a terpene source within a room. Proximity to the source yields the most accurate terpene trace. Simultaneously, an economically efficient evaluation identifies the air return channel as the optimal location for measuring total VOC levels from various sources, as evidenced by consistent median volume readings across all 13 sensors employed in the study.



\vspace{2mm}
\textit{\textbf{Key Takeaway:
The best sensor location is the return air.}} 
\vspace{2mm}

Experiment 3 was designed to answer question 3, focused on comparing VOC traces emitted by plants in normal and stressed conditions, with a specific emphasis on VOC over Time diagram interpretation factors. By subjecting basil plants to controlled conditions within a test box, we aimed to elucidate differences in terpene emissions under varying plant conditions and important factors to consider while reading a VOC-time diagram. 

The observed variations in VOC emissions between normal and stressed conditions provide valuable insights into plant-terpene dynamics. The significant increase in VOC levels, particularly terpenes, under stressed conditions emphasizes the plant's response to environmental stimuli. Factors such as sunlight exposure, temperature, humidity, and time of day were found to influence the magnitude of VOC emissions from the basil plant, highlighting the multifaceted nature of plant-terpene interactions. These findings contribute to our understanding of how environmental factors modulate VOC emissions from live plants and underscore the importance of comprehensive environmental monitoring in indoor settings.

Also, in this experiment, we meticulously examined various factors influencing the monitoring of VOCs, including the VOC baseline, VOC max point, VOC rate, and sensor placement within the testing environment. These findings underscore the complexity of VOC monitoring and the importance of considering multiple factors when interpreting VOC data in indoor environments.

\subsection{Potential Applications}

There would be many different areas we can use this approach to gain benefits of this approach:  

\textit{Air Quality Monitoring and Health}:
This signal analysis method and its subsequent interpretation can be instrumental in comprehending the behaviour of VOCs within indoor environments. Such understanding serves as a valuable resource for their mitigation and finding optimal solutions to enhance indoor air quality for better health.

\textit{Stress Monitoring and Agriculture}:
The release of plant VOCs can provide valuable insights into identifying stress factors affecting plants, understanding their behaviour based on VOC release patterns, and distinguishing differences among these patterns. Cultivating a deeper understanding of plant stress and its continuous monitoring can empower farmers and agricultural practitioners to implement smarter and more informed strategies for plant care and cultivation.

\textit{Virus Detection and biology}:
Furthermore, the initial approach presented in this paper will serve as a foundation for our ultimate objective in this ongoing research, which is the detection of viruses based on plants' VOC responses. Our goal is to identify the type of virus and its concentration using the VOC trace of plants when exposed to a specific virus. Furthermore, by utilizing genetic manipulation and protein modification techniques, we can harness the potential of genetically altered plants to tailor their responses according to our desired outcomes. This allows for targeted adjustments within specific plant-virus interactions.

\textit{Other applications}: We believe that this approach will open doors to a new avenue of research for science enthusiasts. Therefore, we invite them to join us in exploring additional facets of this innovative sensor-plant approach.

\section{Future Work}

This paper aims to function as a building block for this novel research approach. In this context, there will be numerous questions arising, which we plan to address in our ongoing research, based on the results of this study: 

First - More data can be gathered for different spaces and different sensor locations to see how the room size impacts the commercial sensor. Based on the results of the emission study, the sensor location can have an impact on the readings. By understanding this impact, we can modify the simple mass-based equation to better predict the emission rate and terpene in the room. As a result, we can probably find a new physics model based on the signal interpretation(using VOC signal baseline, amplitude, slopes, and area under the signal.)

Second - Different sensors can be used in tandem to gather the data. The 4 Awair sensor used for this study could have their own sensitivity meaning that other sensors may be better. Furthermore, expensive sensors can detect unique responses to different VOCs~\cite{dragonieri2017electronic}. Our machine learning models could get better accuracy if there is less noise in the data and to that end we propose creating a new E-nose by combining multiple lower complex sensors.

Thirdly - Gathering plant data can also be improved. As plant VOCs can be impacted by various factors and can sometimes be inconsistent, it is important to have a stable environment where the plant conditions are maintained and VOCs from the plant are captured. This is vital to getting plants to work as sensors as they need to emit enough VOCs for there to be a signal. To do this, we suggest using an eco-chamber where plant conditions such as light, water, soil, and humidity are maintained. This eco-chamber would take in air from the outside environment. After a set amount of time, the VOCs built up in the eco-chamber would be released into a container holding the commercial sensor so it can capture the VOC signal.

Finally, in a bigger environment with a couple of competing VOC sources by using signal processing techniques like FFT(Fast Fourier Transform), Convolution, and Neural Network we can go deeper into signals' interpretation, their different sources, and the possibility of finding viruses using plants' VOC-releasing reactions. 

\section{Conclusion}
Interpreting raw plant terpene data from commercial sensors is a multifaceted challenge, entangled with intricacies such as sensor sensitivity, airflow dynamics, optimal sensor placement, and the inherent variations among different terpenes. This paper represents a foundational foray into the synergy of commercial sensors and machine learning for unraveling the complexities of plant VOCs. Although our study validates the capability of commercial sensors in detecting plant terpenes, it is essential to acknowledge the current limitations in translating this data into real-world equations. Nevertheless, our research establishes a compelling proof of concept, showcasing the potential of machine learning to discern specific VOCs in the air, detect plant stress, identify optimal sensor placements within a room, and successfully differentiate between two distinct chemicals using readily available sensors. These findings not only contribute significantly to our comprehension of plant terpene detection but also pave the way for future investigations, encouraging the development of more sophisticated test beds to yield accurate and practical plant data in diverse environmental scenarios.

\section*{Authorship Contribution} 
\textbf{Seyed Hamidreza Nabaei}: Conceptualization, Research Design and Experimental Methodology, Data Analysis, Experiment Developing, Investigation, Coding, Writing – original draft;\\
\textbf{Ryan Lenfant}: Conceptualization, Research Design, and Experimental Methodology, Data Analysis, Investigation, Coding, Writing – original draft; \\
\textbf{Viswajith Govinda Rajan}: Investigation, Writing – original draft; \\
\textbf{Dong Chen}: Supervision, Writing \& editing;\\
\textbf{Michael Timko}: Supervision, Writing - review \& editing;\\ 
\textbf{Arsalan Heydariyan}: Supervision, Writing - review \& editing;\\ 
\textbf{Bradford Campbell}: Supervision, Writing - review \& editing. 

\section*{Declaration of Generative AI and AI-assisted technologies in the writing process}
Statement: During the preparation of this work the authors used ChatGPT in order to check the sentence structure and grammar. After using this tool/service, the author(s) reviewed and edited the content as needed and take(s) full responsibility for the content of the publication.

\bibliographystyle{cas-model2-names}
\bibliography{ref}
\end{document}